\documentclass{jfm}
\usepackage{graphicx}
\usepackage{newtxtext}
\usepackage{newtxmath}
\usepackage{natbib}
\usepackage{hyperref}
\usepackage{bm}
\usepackage{caption}
\usepackage[dvipsnames]{xcolor}
\hypersetup{
    colorlinks = true,
    urlcolor   = blue,
    citecolor  = CadetBlue,
}

\title{Turbulent convection in rotating slender cells}

\author{Ambrish Pandey\aff{1,2}
        \corresp{\email{ambrish.pandey@ph.iitr.ac.in}}
        \and Katepalli R. Sreenivasan\aff{2,3}}

\affiliation{\aff{1} Department of Physics, Indian Institute of Technology Roorkee, Roorkee 247667, Uttarakhand, India
\aff{2} Center for Space Science, New York University Abu Dhabi, Abu Dhabi 129188, UAE
\aff{3} Tandon School of Engineering, Department of Physics, and Courant Institute of Mathematical Sciences, New York University, New York, NY 11201, USA
}

\begin{document}
\maketitle

\begin{abstract}
Turbulent convection in the interiors of the Sun and the Earth occurs at high Rayleigh numbers $Ra$, low Prandtl numbers $Pr$, and different levels of rotation rates. To understand the combined effects better, we study rotating turbulent convection for $Pr = 0.021$ (for which some laboratory data corresponding to liquid metals are available), and varying Rossby numbers $Ro$, using direct numerical simulations (DNS) in a slender cylinder of aspect ratio 0.1; this confinement allows us to attain high enough Rayleigh numbers. We are motivated by the earlier finding in the absence of rotation that heat transport at high enough $Ra$ is similar between confined and extended domains. We make comparisons with higher aspect ratio data where possible. We study the effects of rotation on the global transport of heat and momentum as well as flow structures (a) for increasing rotation at a few fixed values of $Ra$ and (b) for increasing $Ra$ (up to $10^{10}$) at the fixed, low Ekman number of $1.45 \times 10^{-6}$. We compare the results with those from unity $Pr$ simulations for the same range of $Ra$ and $Ro$, and with the non-rotating case over the same range of $Ra$ and low $Pr$. We find that the effects of rotation diminish with increasing $Ra$. These results and comparison studies suggest that, for high enough $Ra$, rotation alters convective flows in a similar manner for small and large aspect ratios, and so useful insights on the effects of high thermal forcing on convection can be obtained by considering slender domains.
\end{abstract}


\section{Introduction}
\label{sec:intro}

Convection in most natural settings, such as the Earth's interior and Jupiter's atmosphere ~\citep{Heimpel:Nature2005} and the interior convection of the Sun ~\citep{Hanasoge:ARFM2016}, coexists with rotation. Rotating Rayleigh-B\'enard convection (RRBC), where a fluid layer rotates uniformly about its vertical axis and is simultaneously heated from the bottom and cooled from the top, is a popular model of such flows~\citep{Ecke:ARFM2023}. The characteristics of RRBC depend on the Prandtl number $Pr$ (the ratio of the heat and momentum diffusion time scales), the Rayleigh number $Ra$ (the ratio of the buoyancy force to effects of thermal diffusivity and viscosity of the fluid), and the Ekman number $Ek$ (the time scale ratio of rotation and momentum diffusion). $Pr$ is small in many natural convective flows ($\sim 0.01-0.1$ in Earth's outer core~\citep{Aurnou:PEPI2015, Pandey:JFM2022}, $\sim 10^{-6}$ in Sun's interior~\citep{Schumacher:RMP2020}). Despite the importance of the low-$Pr$ RRBC, and the awareness that it is distinct from convection at moderate and high-$Pr$~\citep{King:PNAS2013, Horn:JFM2017, Aurnou:JFM2018}, it has not been explored as extensively as its high-$Pr$ counterpart. In this paper, we study RRBC at the low-$Pr$ of 0.021 for a range of rotation rates, with the $Ra$-range that includes the onset of convection as well as the turbulent state. 

To optimize computational resources (also see the discussion at the end of \S~2), we use a cylindrical domain of aspect ratio $\Gamma = 0.1$. Here $\Gamma$ is the diameter to the height ratio of the cell. We recently demonstrated \citep{Pandey:EPL2021, Pandey:PD2022} that many properties of convective flows in the slender cell are similar, in the absence of rotation, to those in extended domains of $\Gamma = 25$, when $Ra$ is large. Similarly, while the flow structures near the onset of convection indeed depend on $\Gamma$, they may be expected to be similar between confined and extended domains if $Ra$ is large. In any case, wherever possible, we make explicit comparisons with data from wider convection cells. Note, however, that directional confinement has been observed to alter the flow properties in different ways in RBC depending on the control parameters~\citep{Wagner:PoF2013, Chong:PRL2015, Chong:JFM2016}. For example, \citet{Chong:PRL2015} found for $Pr = 4.38$ that in rectangular domains of dimensions $(H,L_y,H)$ the heat transport gets amplified and attains a maximum when $\Gamma_y = L_y/H$ decreases and reaches a certain $Ra$-dependent critical $\Gamma_y$. The narrower boxes, however, were observed to become increasingly resistant to the momentum transport. On the other hand, \citet{Wagner:PoF2013} observed for $Pr = 0.786$ that both the heat and momentum transports generally decrease when $\Gamma_y$ is changed from 1 to $1/10$. 

For comparison purposes, we also perform the DNS of convection in rapidly rotating and non-rotating cells for the same range of $Ra$, while maintaining $Pr$ low at 0.021. We study the effects of rotation on flow structures as well as global heat and momentum transports.  Specifically, we consider the following: 

(1) The effect of rotation on the critical Rayleigh number, $Ra_c$. We elucidate the change in the large structure of the flow, in particular the evolution of the organized helical structure at low $Ra$ into one with increasing small scale content. 

(2) The effect of rotation near the onset of instability. For horizontally unbounded rotating layers, linear stability theories show that the onset of convection is delayed in $Ra$, with the critical Rayleigh number $Ra_c$ and the corresponding length scale $\ell_c$ depending only on $Ek$ when $Pr$ is moderate and large~\citep{Chandrasekhar:book}. For low $Pr$ as well, the dependence of onset parameters on $Pr$ is explicitly known~\citep{Chandrasekhar:book, Zhang:book2017}. What is not known is the behavior of the heat transport for low $Pr$. For moderate $Pr$, the excess heat transport~\citep{Ecke:PRL2014, Plumley:ESS2019, Kunnen:JT2021} given by $Nu-1$ increases linearly with the supercriticality $\epsilon = Ra/Ra_c-1$ \citep{Gillet:JFM2006, Ecke:PLA2015, Long:JFM2020}---the Nusselt number $Nu$ being the ratio of the actual heat transport to that enabled by conduction alone---but the corresponding behavior of low $Pr$ has not yet been explored.

(3) The scaling of heat and momentum transport for large $Ra$ range. A range of scaling exponents $\beta$ in the empirical relations $Nu \sim Ra^\beta$ has been observed in RRBC. In the rapidly rotating regime, $\beta$ is as large as 3.6 for convection in water, with $\beta$ decreasing as rotation decreases~\citep{King:Nature2009, King:JFM2012, Cheng:GJI2015}. Asymptotic simulations of RRBC have revealed that the heat transport scales as $Nu-1 = Ra^{3/2} Pr^{-1/2} Ek^2$ in the geostrophic regime~\citep{Julien:PRL2012, Aurnou:PRR2020, Kunnen:JT2021}. In rotating liquid gallium ($Pr \approx 0.025$), \citet{King:PNAS2013} reported $\beta$ values varying from 0.1 to 1.2 in the rotationally influenced regime in a cylindrical cell with $\Gamma \approx 2$, while \citet{Aurnou:JFM2018} found $\beta \approx 0.9$ for a similar aspect ratio ($\Gamma = 1.9$). We examine the validity of these expectations.

(4) The bulk temperature gradient in the rotating slender cells. The inhibition of turbulent mixing by rotation is often manifested by the presence of significant vertical temperature gradient $\partial T/\partial z$ in the bulk region. This gradient varies non-monotonically in RRBC~\citep{Cheng:PRF2020, Guzman:PRF2022}, and, for moderate $Pr$, the rapidity of its variation with $Ra$ indicates various flow structures \citep{Julien:GAFD2012}. The low-$Pr$ case has been restricted mostly to moderate $Ra$~\citep{King:PNAS2013, Horn:JFM2017, Aurnou:JFM2018, Guzman:PRF2022} because of numerical and experimental challenges \citep{Pandey:JFM2022}. Here, we quantity $\partial T/\partial z$ in the bulk region in both low- and moderate-$Pr$ convection, carrying out the DNS for high $Ra$, and find that it is qualitatively similar to that in wider cells.

(5) Viscous boundary layer near the horizontal plate. In non-rotating convection, the viscous boundary layer near the plates becomes thinner with increasing thermal forcing, whereas its width $\delta_u$ is determined by the Ekman number in RRBC; in rapidly rotating convective flows, $\delta_u \sim \sqrt{Ek}$~\citep{King:JFM2013}. We estimate $\delta_u$ and find that it scales as $\sqrt{Ek}$ in the rotating slender cells when rotation effects dominate the thermal forcing. We further compare the velocity profile in the near-wall region and observe very good agreement with the analytical Ekman layer profile~\citep{Guzman:PRF2022} in the regime where $\delta_u \sim \sqrt{Ek}$ scaling holds well.

As the onset length scale decreases with decreasing Ekman number, convective structures grow in number with decreasing $Ek$ in a domain of fixed $\Gamma$. This aspect has been utilized by researchers by exploring rotating convection at low $Ek$ (and high $Ra$) in slender convection domains because the effects of confinement may be rendered insignificant by the presence of a multitude of elementary flow structures~\citep{Cheng:GJI2015, Cheng:GAFD2018, Cheng:PRF2020, Madonia:EPL2021}. However, flow properties in confined RRBC domains could be altered in an intricate manner---for example by the so called boundary zonal flow~\citep{Zhang:PRL2020, Shishkina:JFM2020, Zhang:JFM2021, Ecke:PRF2022, Wedi:JFM2022} or sidewall circulation~\citep{deWit:PRF2020, Favier:JFM2020}. In the present work, the slender convection cell contains between 1 and 3 elementary structures at the onset, clearly indicating that the flow is confined in that context. In spite of this, the way the flow is altered due to rotation is essentially the same as in flows in wider cells, especially at higher $Ra$. 

After a brief discussion of the simulation tools in \S\ref{sec:numerical}, we present comments on flow morphology in \S\ref{sec:structure}. Flow structures near the onset of convection are discussed in \S\ref{sec:onset}, the scaling results on global heat transport in \S\ref{sec:transport}, and the temperature gradient in the bulk region and the viscous boundary layer in \S\ref{sec:dTdz}. A few concluding remarks are presented in \S\ref{sec:concl} while important parameters of our simulations are summarized in appendix~\ref{sec:app_sim}.

\section{Simulation methodology}
\label{sec:numerical}

We solve the non-dimensional Oberbeck-Boussinesq equations
\begin{eqnarray}
\frac{\partial {\bm u}}{\partial t} + {\bm u} \cdot \nabla {\bm u} & = & -\nabla p + T \hat{z} - \frac{1}{Ro} \hat{z} \times {\bm u} + \sqrt{\frac{Pr}{Ra}} \nabla^2 {\bm u}, \label{eq:u_n} \\
\frac{\partial T}{\partial t} + {\bm u} \cdot \nabla T & = & \frac{1}{\sqrt{Pr Ra}} \nabla^2 T,
 \label{eq:T_n} \\
\nabla \cdot {\bm u} & = & 0 \label{eq:m_n},
\end{eqnarray}
where ${\bm u} \, ( \equiv u_x \hat{x} + u_y \hat{y} + u_z \hat{z})$, $T$, and $p$ are the velocity, temperature, and pressure fields, respectively. The normalizing length $H$ is the height between the horizontal plates and $\Delta T$ is the temperature difference between them. The free-fall velocity $u_f = \sqrt{\alpha g \Delta T H}$ and the free-fall time $t_f = H/u_f$ are the relevant velocity and time scales. The Rayleigh number $Ra = \alpha g \Delta T H^3/(\nu \kappa)$ and the Prandtl number $Pr = \nu/\kappa$. The convective Rossby number  $Ro = u_f/(2\Omega H) = \alpha g \Delta T/(2\Omega u_f)$ is the ratio of the buoyancy and Coriolis forces, where $\Omega$ is the rotation rate, and $\alpha, \nu, \kappa$ are the isobaric coefficient of thermal expansion, kinematic viscosity, and thermal diffusivity of the fluid, respectively.

The simulations correspond to $Pr = 0.021$ and $2 \times 10^7 \leq Ra \leq 10^{10}$ in a cylindrical cell of $\Gamma = 0.1$ using the solver {\sc Nek5000}, based on the spectral element method~\citep{Fischer:JCP1997}. The no-slip boundary condition is prescribed for the velocity field on all walls, and the isothermal and adiabatic conditions for the temperature field on horizontal and sidewalls, respectively. The cylinder is decomposed into $N_e$ elements and the turbulence fields within each element are expanded using the $N^{\rm th}$-order Lagrangian interpolation polynomials. Thus, the number of mesh cells in the entire flow is $N_e N^3$; higher mesh density in the near-wall regions is used to capture rapid variations of the  field variables. More details can be found in \citet{Scheel:NJP2013, Iyer:PNAS2020, Pandey:PD2022}. (Incidentally, the number of spectral elements $N_e$ in \citet{Iyer:PNAS2020} was 192,000 for $Ra=10^8, 10^9, 10^{10}$ and $10^{11}$.)

\begin{figure}
\captionsetup{width=1\textwidth}
\centerline{\includegraphics[width=\textwidth]{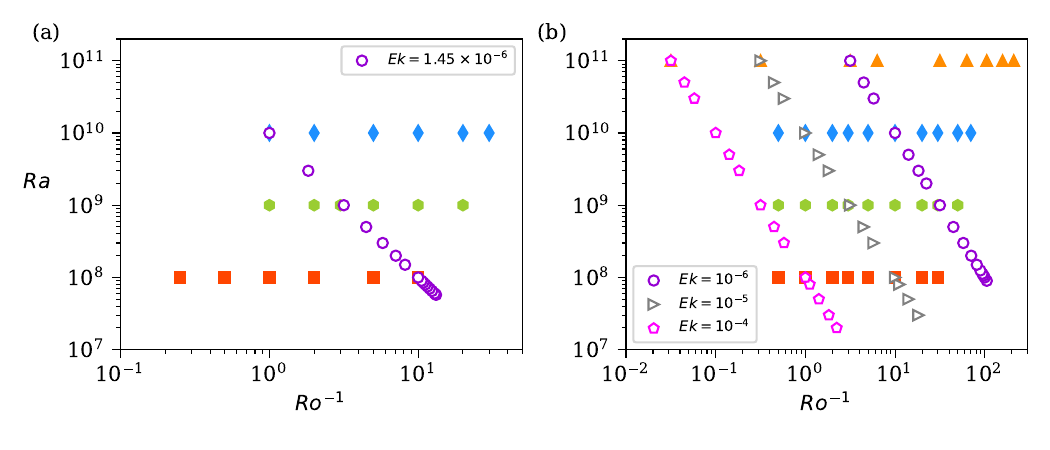}}
\caption{The parameter space explored in the present study for (a) $Pr = 0.021$ and (b) $Pr = 1$. Open symbols are for simulations with fixed rotation and varying thermal forcing, whereas the filled ones are for simulations with fixed forcing and varying rotation rate. In both (a) and (b), the sloping data are for variable $Ro$ but constant $Ek$.}
\label{fig:param}
\end{figure}

The effects of rotation are studied using two different approaches. First, the effects of increasing thermal forcing are explored for a fixed $Ek = 1.45 \times 10^{-6}$ and varying $Ra$ up to $10^{10}$. The Ekman number $Ek = \nu/(2 \Omega H^2)$ quantifies the strength of the viscous force relative to that of the Coriolis force, so we are dealing with a rapidly rotating case. Second, the effects of increasing rotation are studied by fixing $Ra = 10^8, 10^9$, and $10^{10}$ and by decreasing the Rossby number for each $Ra$. Note that the convective Rossby number is also expressed as $Ro = Ek \sqrt{Ra/Pr}$; for a fixed $Ra$ and $Pr$, $Ek$ decreases with the decreasing $Ro$. The simulations for non-rotating convection serve as the reference state. To compare the flow properties with those of moderate-$Pr$ convection, we additionally conduct RRBC simulations for $Pr = 1$ and $Ra$ up to $10^{11}$, but the emphasis in this paper is the low-$Pr$ case. The parameter space in this study is shown in figure~\ref{fig:param}.

The Kolmogorov length scale is estimated as $\eta = (\nu^3/\varepsilon_u)^{1/4}$, where $\varepsilon_u$ is the kinetic energy dissipation rate computed at each point in the flow as
\begin{equation}
\varepsilon_u({\bm x}) = \frac{\nu}{2} \sum_{i,j} \left(\frac{\partial u_i}{\partial x_j} + \frac{\partial u_j}{\partial x_i} \right)^2, 
\end{equation}
where $i,j \equiv (x,y,z)$. To ensure the adequacy of the spatial resolution, we estimate the height-dependent Kolmogorov scale $\eta(z)$ using the area and time averaged dissipation rate $\langle \varepsilon_u \rangle_{A,t}(z)$, and ensure that the vertical grid spacing $\Delta_z$ remains of the order $\eta(z)$. This constraint captures all significant variations in the velocity field. Further, within the Ekman layer, which varies as $\delta_\nu \sim \sqrt{Ek}$, we have embedded 5-20 grid points.

We briefly expand here on the computational gains in using a slender cell because the fluid volume is smaller by a factor of $\Gamma^2$. Higher $Ra$ could thus be achieved for the same computational resources, compared to those of higher $\Gamma$. However, an increased fraction of fluid is affected by the sidewall and the critical $Ra$ for the onset of convection grows for small $\Gamma$~\citep{Shishkina:PRF2021, Ahlers:PRL2022}. To that extent the computational advantage of using a slender domain to explore highly turbulent regime of convection tends to be diminished, but one needs further exploration on these advantages in different Rayleigh number regimes.

\section{Flow morphology}
\label{sec:structure}

\begin{figure}
\captionsetup{width=1\textwidth}
\centerline{\includegraphics[width=\textwidth]{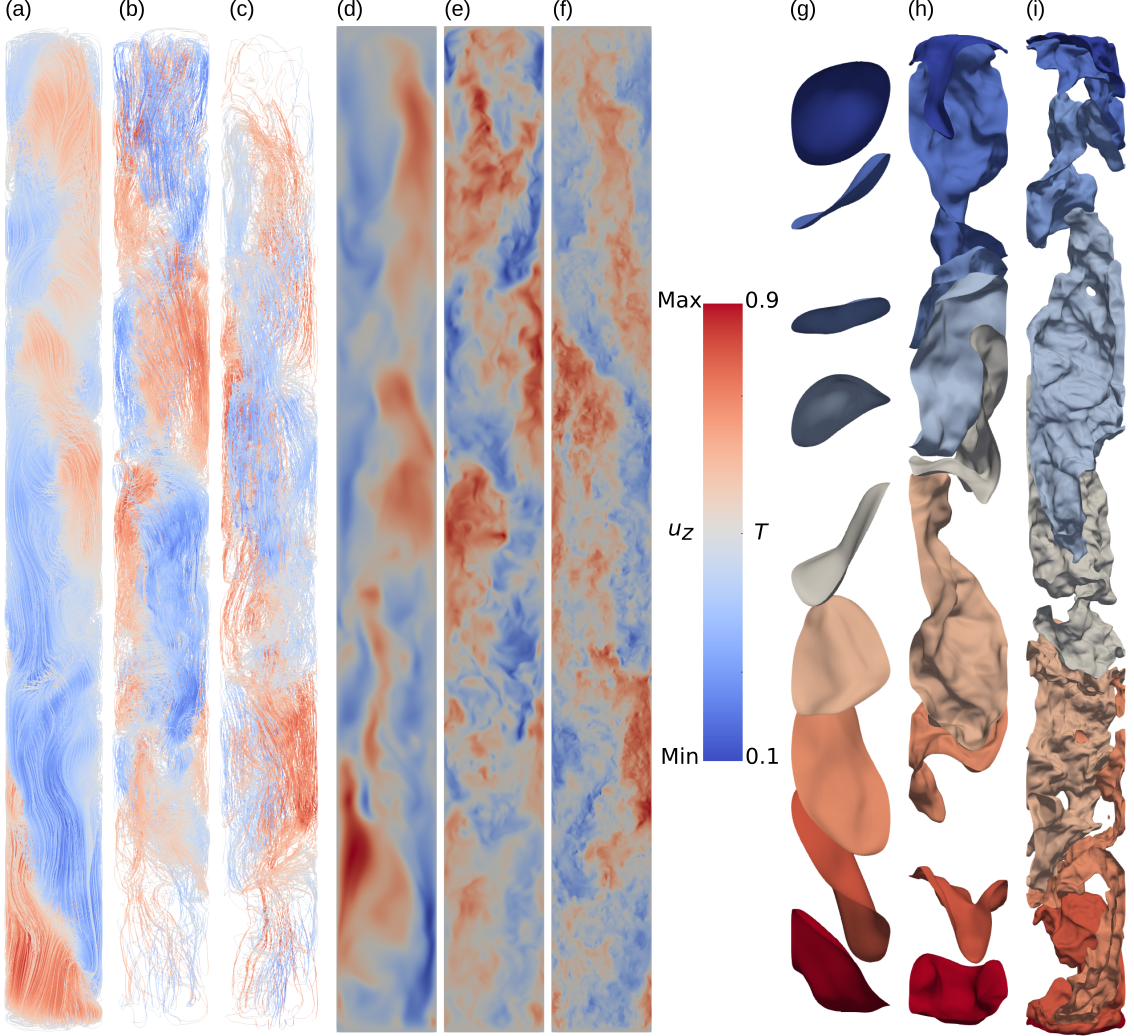}}
\caption{Instantaneous convective structures in a non-rotating slender cell for $Pr = 0.021$ and $Ra = 10^8$ (a,d,g), $Ra = 10^9$ (b,e,h), and $Ra = 10^{10}$ (c,f,i). The velocity streamlines (a--c), colored by the vertical velocity, exhibit helical flow structures in the slender cell. Planar cuts of the vertical velocity (d--f) reveal that progressively finer flow structures are generated with increasing thermal forcing. Isosurfaces of the temperature (g--i) indicate that, despite increased mixing with $Ra$, the isothermal bulk region, observed to exist in wider convection domains, is not present in the slender cell.}
\label{fig:iso_nonrot}
\end{figure}

Multiple vertically-stacked circulation rolls lead to helical structures in slender convection domains~\citep{Iyer:PNAS2020, Zwirner:PRL2020, Pandey:EPL2021, Pandey:PD2022}. The flow configuration in the non-rotating slender cell is shown in figure~\ref{fig:iso_nonrot} for varying $Ra$. The instantaneous velocity streamlines shown in figure~\ref{fig:iso_nonrot}(a--c), coloured according to the vertical velocity, confirm the presence of vertically-stacked rolls. The helical flow structure is relatively smooth for $Ra = 10^8$ (panel (a)) but becomes increasingly complex as the thermal forcing increases. The vertical velocity slices in figure~\ref{fig:iso_nonrot}(d--f) exhibit coherently moving flows, both up and down, with sizes comparable to the lateral extent of the flow. However, these organized structures incorporate increasingly smaller scales as $Ra$ increases. The corresponding temperature isosurfaces in figure~\ref{fig:iso_nonrot}(g--i) show that the mixing is weak at low $Ra$ but becomes increasingly effective as the thermal forcing becomes stronger. Even in a highly turbulent flow for $Ra = 10^{10}$ (figure~\ref{fig:iso_nonrot}(i)), a variety of temperature isosurfaces are present in the bulk region, which indicates that the turbulent mixing is weaker than in wider convection domains, where a well-mixed and isothermal bulk component is observed. The global heat transfer, however, is not very different in the two cases \citep{Pandey:JFM2022}.

The critical parameters for the onset of non-rotating convection are independent of $Pr$. In contrast, the onset parameters in rotating convection do depend on the Prandtl number when it is less than 0.68~\citep{Chandrasekhar:book}. Linear stability analysis for $Pr > 0.68$ in horizontally unconfined domains yield the Rayleigh number and the length scale for the steady onset as
\begin{eqnarray}
Ra_c & = & 3 (\pi^2/2)^{2/3} Ek^{-4/3} \approx 8.7 Ek^{-4/3}, \label{eq:Rac_st} \\
\ell_c/H & = & (2\pi^4)^{1/6} Ek^{1/3} \approx  2.4Ek^{1/3}. \label{eq:lc_st}
\end{eqnarray}
For low Prandtl numbers ($Pr < 0.68$), the critical parameters at the oscillatory onset depend on both $Ek$ and $Pr$ ~\citep{Horn:JFM2017, Aurnou:JFM2018, Vogt:JFM2021} as
\begin{eqnarray}
Ra_c & = & 3 \pi \left( \frac{2 \pi}{1+Pr} \right)^{1/3} \left( \frac{Ek}{Pr} \right)^{-4/3}, \label{eq:Rac_osc} \\
\ell_c/H & = & (2 \pi^4)^{1/6} (1+Pr)^{1/3} \left( \frac{Ek}{Pr} \right)^{1/3}, \label{eq:lc_osc} \\
\omega_c & = & (2 - 3Pr^2)^{1/2} \left( \frac{2 \pi}{1+Pr} \right)^{2/3}  \left( \frac{Ek}{Pr} \right)^{1/3}. \label{eq:omc_osc}
\end{eqnarray}
Thus, the onset length scale $\ell_c$ in low-$Pr$ convection is larger by a factor of $(1+1/Pr)^{1/3}$.

\begin{figure}
\captionsetup{width=1\textwidth}
\centerline{\includegraphics[width=\textwidth]{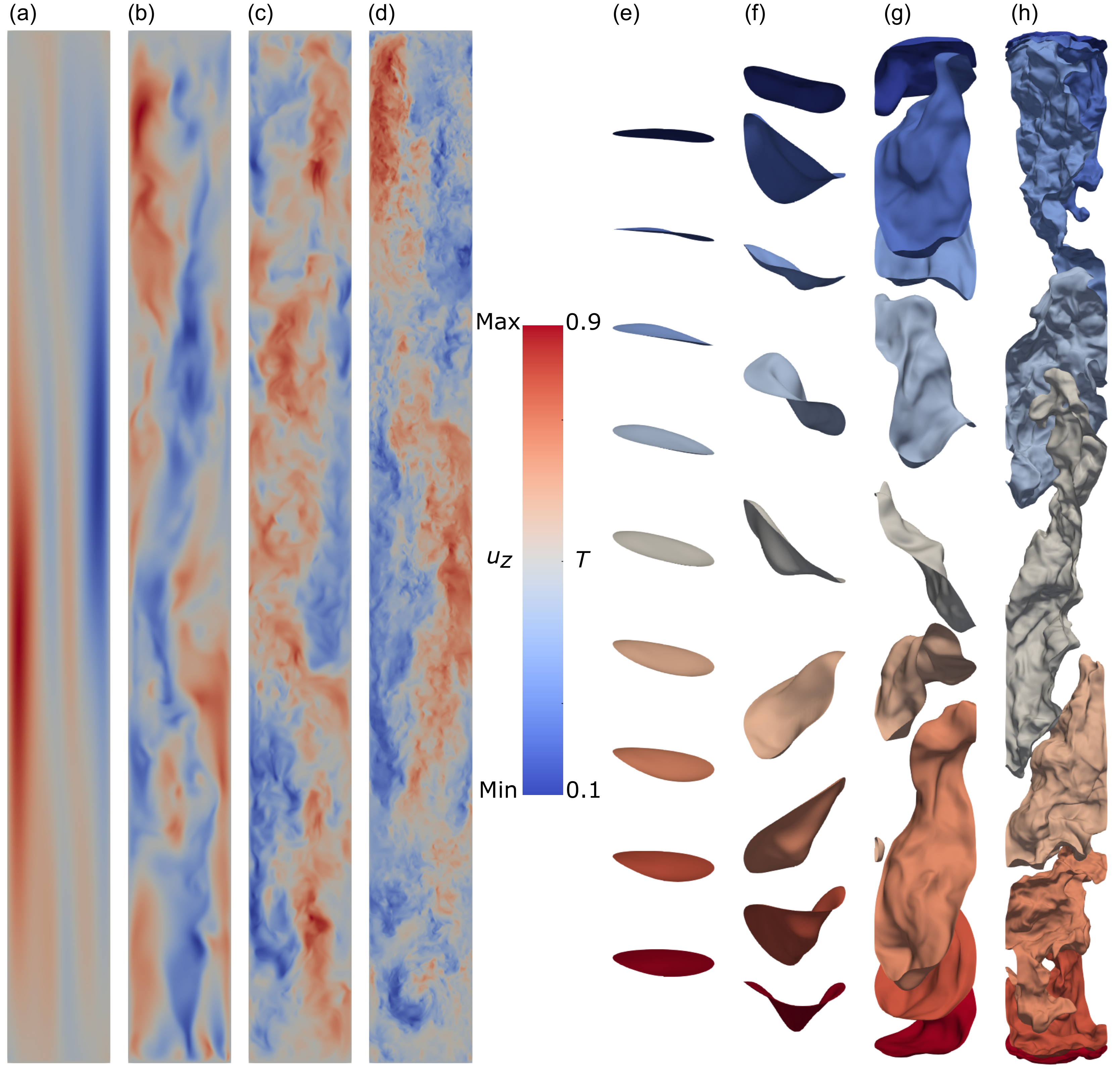}}
\caption{Flow morphology in a rotating slender cell for $Pr = 0.021, Ek = 1.45 \times 10^{-6}$ revealed by instantaneous vertical velocity slices (a--d) and temperature isosurfaces (e--h), for $Ra = 6 \times 10^7$ (a,e), $Ra = 2 \times 10^8$ (b,f), $Ra = 10^9$ (c,g), and $Ra = 10^{10}$ (d,h). Near the onset of convection (panels a,e), flow structures feel the rotation strongly, and the variation along the vertical direction is almost suppressed. With increasing $Ra$, the resilience increases and the flow configuration for $Ra = 10^{10}$ (panels d,h) shows strong resemblance with its non-rotating counterpart in figure~\ref{fig:iso_nonrot} (the global heat and momentum transports are also nearly indifferent for these cases, see table~\ref{table:details_Ra1e10}.)}
\label{fig:iso_Ek_fixed}
\end{figure}

The flow in the rotating cell for $Ek = 1.45 \times 10^{-6}$ is shown in figure~\ref{fig:iso_Ek_fixed} for various $Ra$. For this $Ek$ and $Pr = 0.021$, the length scale at the onset of convection according to~\eqref{eq:lc_osc} is $\ell_c/H \approx 0.1$, which is equal to the horizontal dimension of the slender domain. Thus, the low-$Pr$ flow for this $Ek$ is confined at the onset. For $Ra = 6 \times 10^7$---not far from the onset---the Coriolis force dominates the buoyancy force leading to smooth and tall velocity structures inhabiting the entire depth (figure~\ref{fig:iso_Ek_fixed}(a)). For $Ra = 2 \times 10^8$ (panel b), buoyancy becomes stronger but the flow continues to be influenced by the strong rotation. The observed tall structures develop wavy character as in high-$Pr$ convection~\citep{Cheng:PRF2020}. The vertical coherence is lost nearly completely for $Ra = 10^9$ and, for $Ra = 10^{10}$, the flow morphology appears very close to that in the non-rotating cell shown in figure~\ref{fig:iso_nonrot}(f,i), indicating that the effects of the Coriolis force (for this rotation) essentially vanish around $Ra = 10^{10}$. From the velocity streamlines visualization (not shown), we infer that the helical structure, present for the entire range of the thermal forcing explored in the non-rotating cell, is not observed in the rotating slender convection when the Coriolis force dominates; the helical configuration is recovered only when the thermal forcing becomes strong enough to overcome rotation.

Dwindling vertical coherence with increasing $Ra$, for a fixed rotation, is also clear from the temperature field in figure~\ref{fig:iso_Ek_fixed}(e--h). The temperature isosurfaces for $Ra = 6 \times 10^7$---shown in figure~\ref{fig:iso_Ek_fixed}(e)---are nearly flat circular discs. This is in line with the Taylor-Proudman constraint that the vertical variation of the flow is inhibited in a rapidly rotating inviscid flow~\citep{Chandrasekhar:book}. With increasing $Ra$, the isosurfaces become increasingly three-dimensional and, for $Ra = 10^{10}$, appear very similar to the non-rotating case. In \S~\ref{sec:transport}, we also show that the integral transport properties of the rotating flow at $Ra = 10^{10}$ and $Ek = 1.45 \times 10^{-6}$ are nearly the same as those of the corresponding non-rotating flow.

\begin{figure}
\captionsetup{width=1\textwidth}
\centerline{\includegraphics[width=\textwidth]{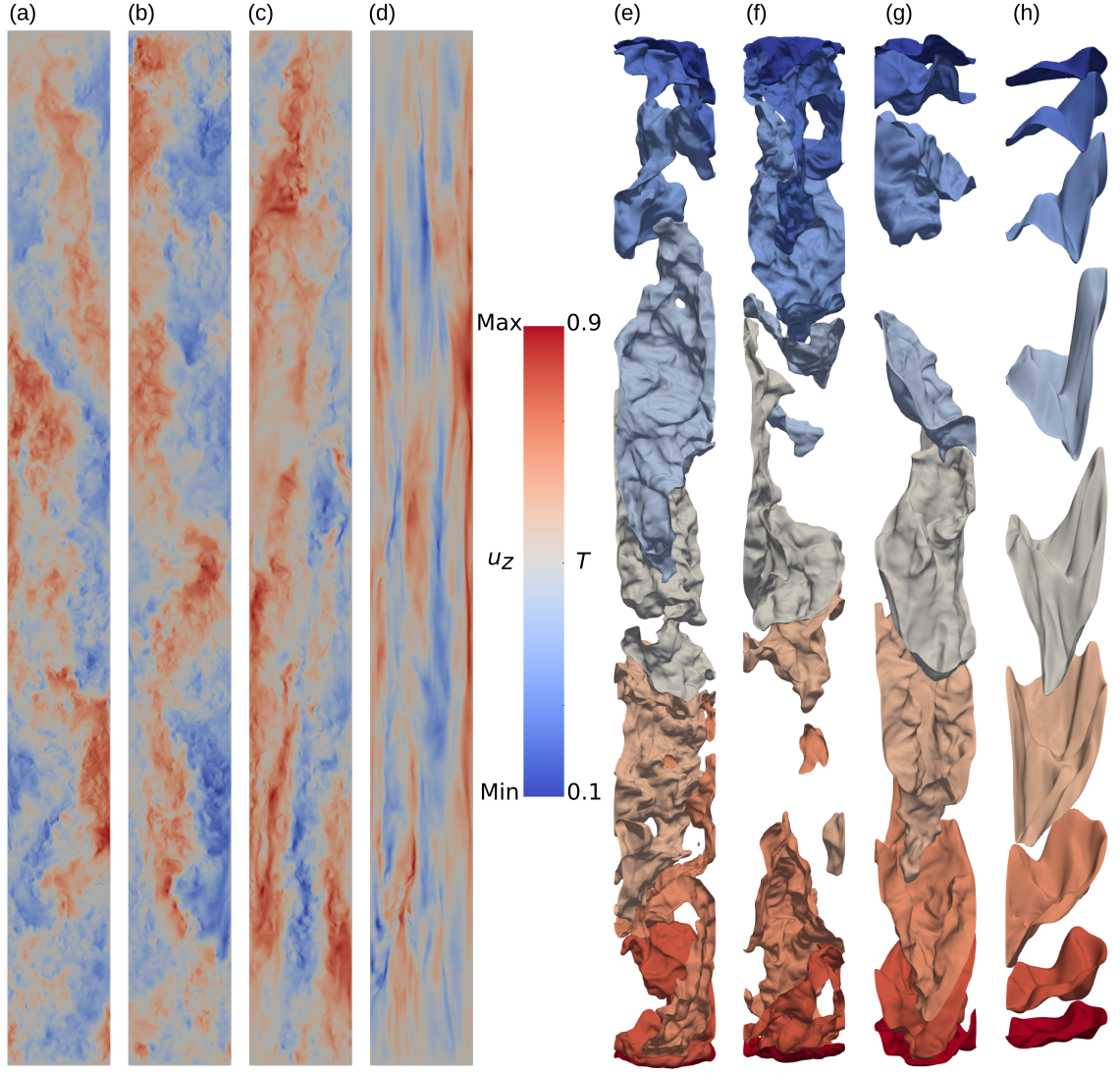}}
\caption{Evolution of the convective structures with increasing rotation rate for $Pr = 0.021, Ra = 10^{10}$: (a,e) $Ro^{-1} = 0$, (b,f) $Ro^{-1} = 2$, (c,g) $Ro^{-1} = 10$, (d,h) $Ro^{-1} = 30$. The flow loses its three-dimensional character, and the length scale of the velocity structures decreases, as the Rossby number decreases.}
\label{fig:iso_Ra1e10}
\end{figure}

A qualitatively similar change in the flow morphology is observed when the rotation increases for a prescribed thermal forcing~\citep{Horn:JFM2015, Aurnou:PRR2020}. Figure~\ref{fig:iso_Ra1e10} exhibits flow structures for $Ra = 10^{10}$ and $0 \leq Ro^{-1} \leq 30$, where the helical structure transforms to tall vertically-elongated velocity structures as the container is rotated increasingly rapidly. The temperature contours also lose their three-dimensional character as $Ro$ decreases, consistent with the observations in wider convection domains filled with moderate and high-$Pr$ fluids~\citep{Cheng:GJI2015}. We also observe from figure~\ref{fig:iso_Ra1e10} that the flow length scale varies with varying $Ro$. For $Ro^{-1} = 30$ in figure~\ref{fig:iso_Ra1e10}(d), $Ek \approx 4.8 \times 10^{-8}$ and the linear stability theory yields $\ell_c/H \approx 0.032$, which is nearly three times smaller than that for $Ek \approx 1.45 \times 10^{-6}$ in figure~\ref{fig:iso_Ek_fixed}. Therefore, the confinement effects in the slender cell are progressively mitigated as $Ek$ decreases.

\section{Rotating slender convection near the onset}
\label{sec:onset}

We first explore the flow evolution in the low-supercritical regime for $Ek = 1.45 \times 10^{-6}$ and $Pr = 0.021$.
We start simulations from the conduction solution with random perturbations and observe that the convective state, corresponding to substantially non-zero values of $Nu-1$, occurs first at $Ra = 5.6 \times 10^7$. Note that this value is nearly an order of magnitude larger than the $Ra_c$ obtained from~\eqref{eq:Rac_osc}.
Here, the Nusselt number $Nu$ is computed as
\begin{equation}
Nu = 1 + \sqrt{Ra Pr} \, \langle u_z T \rangle_{V,t} \,,	 \label{eq:Nu_uzT}
\end{equation}
where $\langle \cdot \rangle_{V,t}$ denotes averaging over the entire flow and integration time. For a simulation at $Ra = 5.55 \times 10^7$ started from the conduction state, we observe $Nu - 1 \approx 0.0022$, whereas we get $Nu -1 \approx 0.049$ when the same simulation is started with a flow state given by the simulation at $Ra = 6 \times 10^7$. By decreasing $Ra$, we can observe finite amplitude convection up to $Ra = 5.40 \times 10^7$, where the convective flux $Nu - 1 \approx 0.034$, small but significantly different from zero. Thus, there is modest hysteresis in low-$Pr$ RRBC in slender cell.

\begin{figure}
\captionsetup{width=1\textwidth}
\centerline{\includegraphics[width=\textwidth]{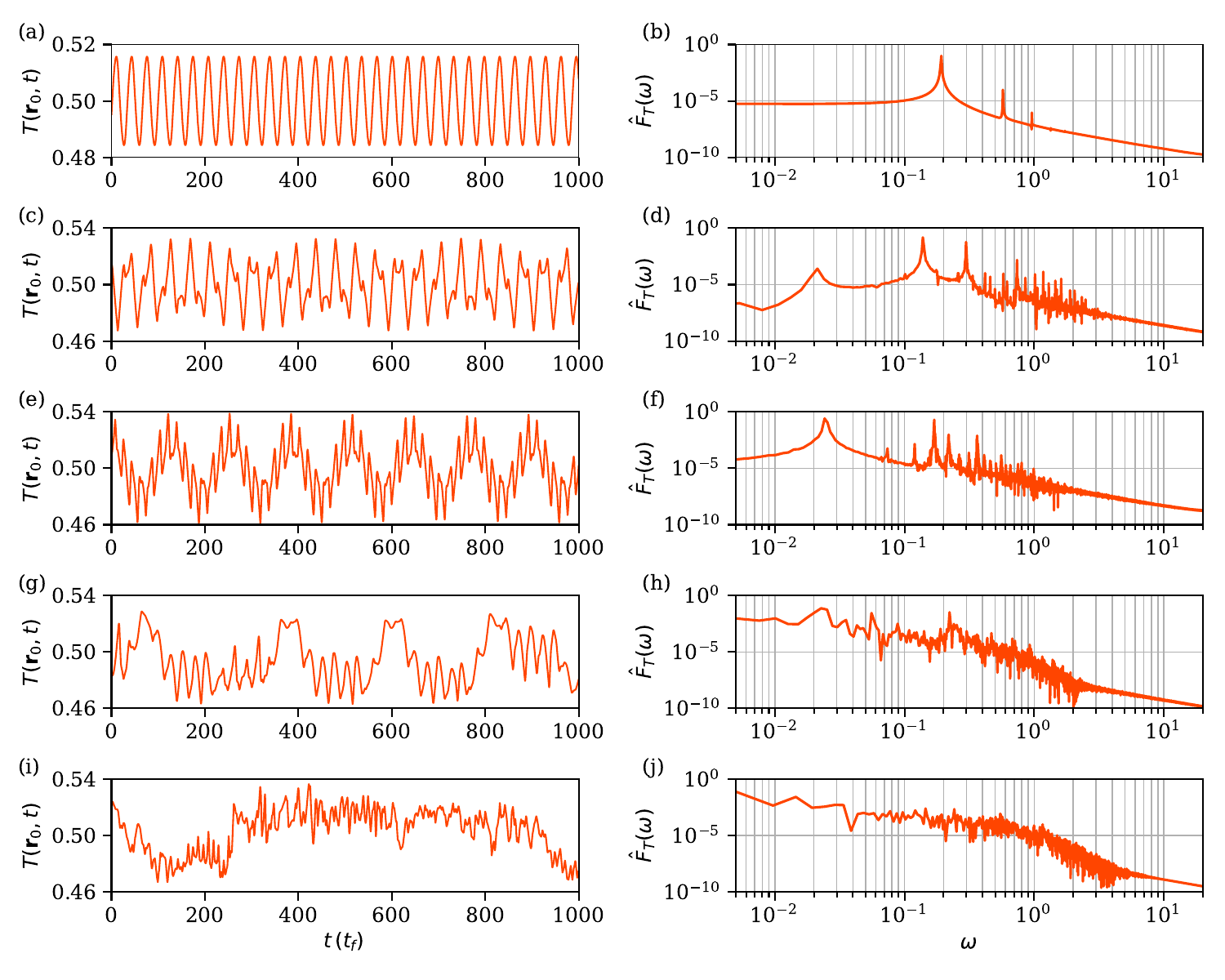}}
\caption{Temperature signal (left column) in the midplane at a probe near the sidewall and the corresponding power spectrum (right column) in a rapidly rotating flow ($Pr = 0.021, Ek = 1.45 \times 10^{-6}$) near the onset of convection: (a,b) $Ra = 6 \times 10^7$, (c,d) $Ra = 7 \times 10^7$, (e,f) $Ra = 8 \times 10^7$, (g,h) $Ra = 9 \times 10^7$, and (i,j) $Ra =  10^8$.}
\label{fig:signal}
\end{figure}

We monitor the evolution of the temperature and velocity fields at a few locations in the flow and show the temperature variation at mid-height near the sidewall in figure~\ref{fig:signal} for $Ra \leq 10^8$ and $Ek = 1.45 \times 10^{-6}$. Figure~\ref{fig:signal}(a) exhibits that the flow evolves periodically for $Ra = 6 \times 10^7$, a feature also observed for lower $Ra$ simulations. The corresponding power spectrum, shown in figure~\ref{fig:signal}(b), reveals a single dominant frequency at $\omega \approx 0.20$ and its higher harmonics. It is interesting that this frequency agrees well with $\omega_c \approx 0.195$ predicted from \eqref{eq:omc_osc} using the linear stability analysis at $Ek = 1.45 \times 10^{-6}$~\citep{Chandrasekhar:book}. With increasing $Ra$ the flow evolution becomes progressively complex due to the emergence of other modes. For $Ra = 7 \times 10^7$, a high-amplitude peak develops also at a lower frequency, which indicates the presence of the wall modes~\citep{Goldstein:JFM1994, Horn:JFM2017, Aurnou:JFM2018}. For $Ra = 8 \times 10^7$, the peak at lower frequency becomes comparably strong to its higher frequency counterpart. The periodicity is nearly lost at $Ra = 9 \times 10^7$ and the flow becomes chaotic. The broadband power spectrum for $Ra \geq 9 \times 10^7$ indicates the presence of flow structures of a wide range of temporal (and spatial) scales. Thus, due to its highly inertial nature, low-$Pr$ RRBC becomes promptly complex. 

\begin{figure}
\captionsetup{width=1\textwidth}
\centerline{\includegraphics[width=\textwidth]{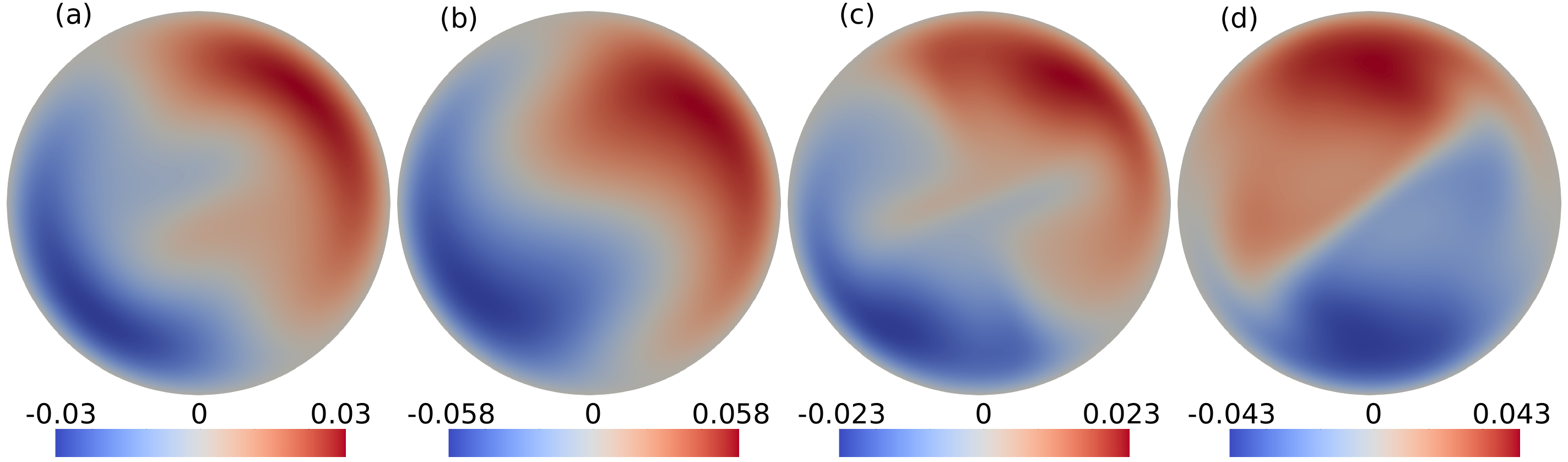}}
\caption{Instantaneous vertical velocity contours in the midplane for $Ek = 1.45 \times 10^{-6}$ and (a) $Ra = 6 \times 10^7$, (b) $Ra = 7 \times 10^7$, (c) $Ra = 8 \times 10^7$, (d) $Ra = 9 \times 10^7$. Peak amplitudes in the velocity are observed near the sidewall at low $Ra$ but the interior of the domain is filled with stronger flows as thermal driving becomes stronger.}
\label{fig:uz_mid}
\end{figure}

Figure~\ref{fig:uz_mid} shows the instantaneous midplane slices of the vertical velocity for $Ra \leq 9 \times 10^7$ in the rotating cell at $Ek = 1.45 \times 10^{-6}$. We can see that the vertical velocity peaks near the sidewall at $Ra = 6 \times 10^7$, while the bulk region (away from the sidewall) is characterized by low amplitude structures. This is a signature of the wall modes in the slender convection cell at a low Prandtl number~\citep{Horn:JFM2017, Aurnou:JFM2018}. For $Ra = 7 \times 10^7$ and $8 \times 10^7$ the high-amplitude patch broadens and encroaches the bulk interior. However, the interior is nearly entirely occupied by the bulk mode at $Ra = 9 \times 10^7$, and has taken over the wall modes~\citep{Goldstein:JFM1994}. Further, the convective flow patterns in rotating cylinders are observed to precess (mostly) in the retrograde direction~\citep{Zhong:JFM1993, Horn:JFM2017}. Similar precessing patterns in the slender cell at low Prandtl number can be found in the supplementary movies for a few cases.

\begin{figure}
\captionsetup{width=1\textwidth}
\centerline{\includegraphics[width=1\textwidth]{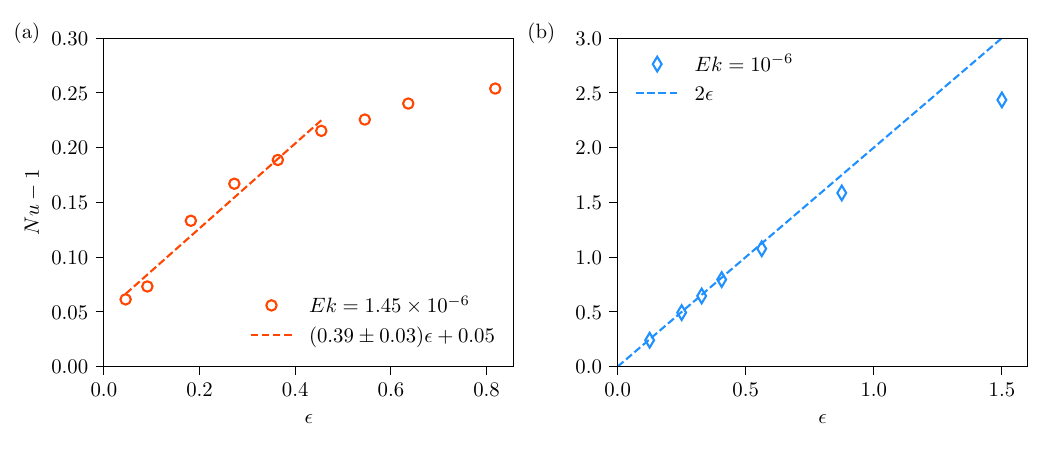}}
\caption{Convective heat transport $Nu-1$ as a function of the normalized distance $\epsilon = Ra/Ra_c-1$ from the onset for (a) $Pr = 0.021$ and (b) $Pr = 1$. Linear scaling is observed in the vicinity of the onset for both the cases but a finite intercept in panel (a) is due to highly inertial nature of low-$Pr$ convection.}
\label{fig:Nu_onset}
\end{figure}

We now compare the heat transport at the onset in flows at low $Pr$ with those at moderate and high $Pr$, both rotating. For moderate and high $Pr$, the convective heat transport $Nu-1$ has been observed to increase linearly with the supercriticality $\epsilon = Ra/Ra_c - 1$~\citep{Ecke:PLA2015, Gillet:JFM2006, Gastine:JFM2016, Long:JFM2020, Ecke:PRF2022}. Figure~\ref{fig:Nu_onset}(a) shows the present data on $Nu-1$ as a function of $\epsilon$ on a linear-linear scale for $Pr = 0.021$ and $Ek = 1.45 \times 10^{-6}$. Even though there is modest hysteresis (as mentioned earlier), we have taken $Ra_c = 5.5 \times 10^7$ based on the observation that the convective heat transport is very small at $Ra = 5.55 \times 10^7$. Figure~\ref{fig:Nu_onset}(a) shows a linear trend for $\epsilon \lesssim 0.5$, with the best fit given by $Nu-1 = 0.39\epsilon + 0.05$. The precise value of the finite intercept depends on the modest hysteresis just mentioned, and so probably not entirely reliable.

The data for unity Prandtl number in the same slender cell at a similar Ekman number, i.e., $Ek = 10^{-6}$ is shown in figure~\ref{fig:Nu_onset}(b). For this case, the heat transport due to convective motion vanishes at $Ra \approx 8 \times 10^7$, this being the onset Rayleigh number. The data follow the linear scaling quite well; when extrapolated back to $Nu =1$, one obtains $Ra_c = 8 \times 10^7$, in perfect agreement with $Ra_c$ determined from inspecting the DNS. It is intriguing that~\eqref{eq:Rac_st} yields $Ra_c \approx 8.7 \times 10^8$, which is an order of magnitude higher than the $Ra_c$ determined from DNS data. This is due to wall modes that lower the critical Rayleigh number in confined domains~\citep{Herrmann:JFM1993, Aurnou:JFM2018, Vogt:JFM2021}. Figure~\ref{fig:Nu_onset}(b) further shows that the prefactor of the linear scaling is $\approx 2$, which is close to 1.54 reported recently in a $\Gamma = 1/2$ cell for $Pr = 0.8$ and $Ek = 10^{-6}$~\citep{Ecke:PRF2022, Ecke:PLA2015}. Thus, a slightly different convective heat flux near the onset could be due to the highly inertial nature of low-$Pr$ convection, where the chaotic time dependence is ingrained even at the onset. It appears fair to conclude, overall, that the onset behavior is essentially the same for all Prandtl numbers.

\section{Global transport of heat and momentum in the turbulent state}
\label{sec:transport}

\begin{figure}
\captionsetup{width=1\textwidth}
\centerline{\includegraphics[width=0.75\textwidth]{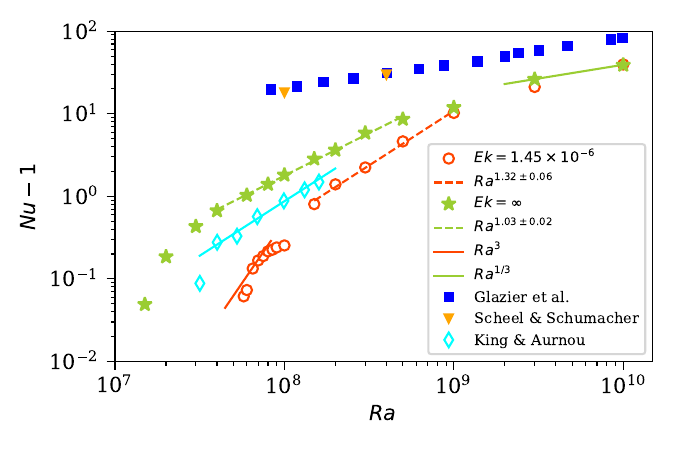}}
\caption{Convective heat transport as a function of $Ra$ in the non-rotating slender cell (green stars) and in a rapidly rotating slender cell (red circles) of $\Gamma = 0.1$ for $Pr = 0.021$. Heat flux in the non-rotating cell exhibits a steeper scaling $Nu-1 \sim Ra^{1.03}$ (dashed green line) compared to those observed in wider convection cells for moderate Rayleigh numbers, but a similar $Ra^{1/3}$ scaling for large Rayleigh numbers (solid green line). $Nu$ in rotating convection is lower than in non-rotating convection when Rayleigh numbers are small but the differences essentially diminish as the thermal forcing increases. The data for $10^8 < Ra \leq 10^9$ exhibits a power law and the best fit yields $Nu-1 \sim Ra^{1.32}$ (dashed red line), which is close to $Nu-1 \sim Ra^{3/2}$ scaling in the geostrophic regime. Cyan diamonds represent experimental data for $Ek = 10^{-6}$ in $\Gamma = 1$ cylinder from \citet{King:PNAS2013} and solid cyan line indicates $Ra^{1.32}$ scaling. Solid lines are not the best fits but are drawn as a guide to the eye. Filled symbols correspond to low-$Pr$ non-rotating convection from the literature: blue squares represent the experimental data from \citet{Glazier:Nature1999} in $\Gamma = 1/2$ domain, whereas orange down-triangles correspond to DNS data in $\Gamma = 1$ cell by \citet{Scheel:PRF2017}.
}
\label{fig:Nu_Ra_rot}
\end{figure}

We now compare the heat transport over an extended range of $Ra$ between rotating and non-rotating cases, both at the low $Pr = 0.021$; see figure~\ref{fig:Nu_Ra_rot}. The data for the non-rotating slender cell (green stars) do not follow a satisfactory power law, but we proceed to fit power laws for different segments of $Ra$ and comment on them. Let us first note that the critical Rayleigh number for the onset of convection in the slender cell is nearly $1.1 \times 10^7$~\citep{Pandey:EPL2021}, which is much higher than that in unbounded domains~\citep{Chandrasekhar:book}. However, the temperature and velocity evolution in the flow as well as the averaged heat flux at the horizontal plates exhibit chaotic time dependence already for $Ra = 2 \times 10^7$. This indicates that the transition to turbulence in the slender cell for $Pr = 0.021$ occurs not far from the onset $Ra$, which is in line with the observations in wider domains~\citep{Schumacher:PNAS2015, Horn:JFM2017}. Figure~\ref{fig:Nu_Ra_rot} also plots the heat transport from non-rotating convection experiments by \citet{Glazier:Nature1999} in a $\Gamma = 1/2$ cell and from DNS by \citet{Scheel:PRF2017} in a $\Gamma = 1$ cell, both at $Pr \approx 0.021$. While the heat transport in the slender cell is lower than those reported in wider cells, the discrepancy decreases with increasing $Ra$; the slender data at the largest $Ra$ explored in this work follows a scaling similar to that in wider convection cells.

It is well known that rotation reduces heat transport~\citep{Chandrasekhar:book, Plumley:ESS2019, Kunnen:JT2021, Ecke:ARFM2023}. The data for $Ek = 1.45 \times 10^{-6}$ (figure~\ref{fig:Nu_Ra_rot} -- red circles) confirm this behavior---except for large $Ra \geq 10^9$, for which the Nusselt numbers in rotating and non-rotating cases are quite close; for these particular conditions, $Nu$ can be said to be essentially unaffected by rotation and the data nearly follow the canonical non-rotating $Ra^{1/3}$ scaling~\citep{Niemela:Nature2000}--- indicating that the effects of rotation on the low-$Pr$ convection in the slender cell resemble those in wider domains. 

Turning attention to low $Ra$ for, say $Ra < 10^8$, $Nu$ is seen to follow a much steeper scaling of $Nu-1 \sim Ra^3$ (solid red line). (This does not contradict the linear scaling shown in figure~\ref{fig:Nu_onset}(a), as these plots use different quantities.) This scaling regime is similar to that reported in DNS in a horizontally-periodic box for $Pr = 1$~\citep{Song:JFM2024}.
Note that a steep heat transport scaling $Nu \sim Ra^3$ near the onset of rotating convection has been proposed by \citet{King:JFM2012}, and so reported for moderate Prandtl numbers~\citep{Stellmach:PRL2014, Cheng:GJI2015}. However, our $Nu$ vs $Ra$ plot (not shown) does not show this cubic scaling near the onset.

In the intermediate region $10^8 < Ra < 10^9$, the data for the rotating case seem to follow a power law with the best fit given by $Nu-1 \sim Ra^{1.32 \pm 0.06}$. This scaling 
is roughly consistent with simulations of the asymptotically reduced equations---describing RBC in the rapidly rotating limit---for which $Nu-1$ increases as $Ra^{3/2}$, for $Pr \geq 0.3$, in the geostrophic regime~\citep{Julien:PRL2012}. A plot of $Nu$ vs $Ra$ (not shown) gives a lower exponent of 0.95 for the same range of $Ra$, which is similar to $Nu \sim Ra^{0.91}$ observed in the ``rotationally-dominated" regime of convection in liquid gallium in a $\Gamma = 1.94$ cylinder~\citep{Aurnou:JFM2018}. Moreover, $Nu-1 \sim Ra^{1.03}$ for intermediate Rayleigh numbers in the non-rotating case. For comparison, we also include data from \citet{King:PNAS2013}, who performed experiments in a $\Gamma = 1$ cylindrical cell for $Pr \approx 0.025$: cyan diamonds in figure~\ref{fig:Nu_Ra_rot} show the heat transport for $Ek = 10^{-6}$. It is clear that $Nu-1$ in the wider RRBC cell also increases steeply near the onset, but for higher $Ra$, $Nu-1$ exhibits a similar $Ra^{1.32}$ scaling (solid cyan line) as observed in the rotating slender cell.

\begin{figure}
\captionsetup{width=1\textwidth}
\centerline{\includegraphics[width=0.75\textwidth]{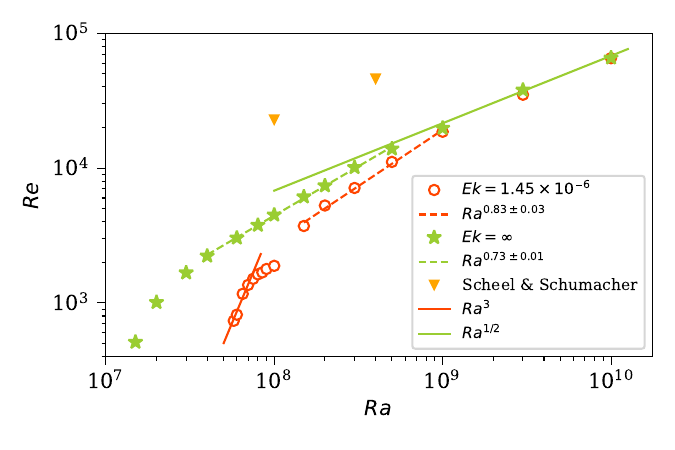}}
\caption{Reynolds number as a function of $Ra$ in the non-rotating cell (green stars) and rotating cell at $Ek = 1.45 \times 10^{-6}$ (red circles). Velocity fluctuations grow rapidly near the onset of convection in the slender cell but the growth rate becomes slower as the driving becomes stronger. The solid green line indicates that the data at the highest $Ra$ nearly follow $\sqrt{Ra}$ power law as in wider cells. Solid red line suggests that $Re$ grows as $Ra^3$ for $Ra < 10^8$. Dashed lines represent the best fits for moderate thermal forcings. The difference between the non-rotating and rotating $Re$ values declines as $Ra$ increases and the two are nearly indistinguishable at $Ra = 10^{10}$. Orange down-triangles represent DNS data in $\Gamma = 1$ cell by \citet{Scheel:PRF2017}. }
\label{fig:Re_Ra}
\end{figure}

Rotation also influences momentum transport, as seen by the behavior of the Reynolds number, based here on the root-mean-square velocity and the depth of the fluid layer, as 
\begin{equation}
Re = \sqrt{\langle u_x^2+u_y^2+u_z^2 \rangle_{V,t} \, Ra/Pr}.
\end{equation}
The Reynolds number in both the non-rotating and rotating cells is plotted as a function of $Ra$ in figure~\ref{fig:Re_Ra}. $Re$ for the non-rotating cell (green stars) increases rapidly near the onset but the rate of increase decays as the thermal driving becomes stronger. In the same intermediate range of $Ra$ where we observe $Nu-1 \sim Ra$ scaling, the best fit yields $Re \sim Ra^{0.73 \pm 0.01}$ scaling. Note that the Reynolds number in wider convection domains has been known to increase nearly as $\sqrt{Ra}$ for moderate and low Prandtl numbers~\citep{Ahlers:RMP2009, Chilla:EPJE2012, Verma:NJP2017, Pandey:POF2016, Scheel:PRF2017, Pandey:JFM2022}. We indicate the $Ra^{1/2}$ scaling by a solid green line in figure~\ref{fig:Re_Ra} and note that the non-rotating data for the largest few Rayleigh numbers of this study nearly follow this same scaling, signalling that the effects of confinement becomes weaker with increasing thermal driving. Also included in figure~\ref{fig:Re_Ra} for comparison are $Re$ computed in a $\Gamma = 1$ cylindrical cell for $Pr = 0.021$ by \citet{Scheel:PRF2017}. We observe that the Reynolds number in the slender cell is smaller compared to that in the wider cell, which is due to a larger effective friction of rigid boundaries in the former case~\citep{Pandey:EPL2021}.

Red circles in figure~\ref{fig:Re_Ra} represent the Reynolds numbers in the rotating slender cell for $Ek = 1.45 \times 10^{-6}$ and the reduced transport of momentum in the presence of rotation is clear~\citep{Schmitz:GAFD2010} for low $Ra$. The figure also shows, similar to figure~\ref{fig:Nu_Ra_rot}, that the onset of convection shifts to higher $Ra$ compared to that for the non-rotating cell. Near the convective onset, $Re$ in the rotating cell approximately grows as $Ra^3$ (solid red line), very similar to the growth of $Nu-1$ in this regime. A rapid growth of $Re$ near the onset of rotating convection was also reported by \citet{Schmitz:GAFD2010}, who performed DNS in a horizontally-periodic domain. With increase of the thermal driving, the growth rate of $Re$ decays and the best fit for $10^8 < Ra \leq 10^9$ is $Re \sim Ra^{0.83 \pm 0.03}$ scaling (dashed red line). This scaling has some similarity with the dissipation-free scaling $Re \sim Ra Ek/Pr$ reported by \citet{Guervilly:Nature2019, Maffei:JFM2021, Vogt:JFM2021, Ecke:ARFM2023}.  For higher $Ra$, $Re$ in the rotating cell approaches that in the non-rotating cell and the difference becomes very small for $Ra > 10^9$. 

\begin{figure}
\captionsetup{width=1\textwidth}
\centerline{\includegraphics[width=\textwidth]{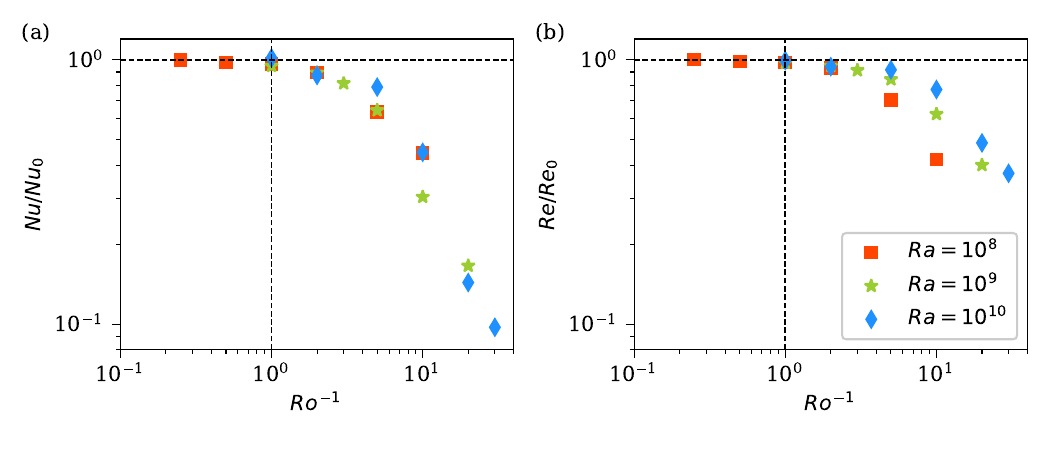}}
\caption{Heat (a) and momentum (b) transports in rotating slender cells normalized with the corresponding values from the non-rotating cell are nearly unity for $Ro^{-1} \leq 2$ but decay rapidly for larger inverse Rossby numbers. The suppression of the heat flux is stronger than that of the momentum flux in low-$Pr$ slender convection.}
\label{fig:Nu_Re_iRo}
\end{figure}

The influence of rotation can be studied also by decreasing the Rossby number $Ro$ for a fixed Rayleigh number~\citep{Stevens:EJMB2013, Kunnen:JFM2011, Ecke:PRL2014, Horn:JFM2015, Aurnou:PRR2020}; the inverse Rossby number $Ro^{-1}$ is a measure of the strength of the Coriolis force relative to buoyancy. We carry out low-$Pr$ simulations for $Ro^{-1} \in [0,30]$ at $Ra = 10^8, 10^9, 10^{10}$. The Nusselt number normalized with $Nu_0$---the heat transport in absence of rotation---as a function of $Ro^{-1}$ is shown in figure~\ref{fig:Nu_Re_iRo}(a), with the curves for different $Ra$ collapsing reasonably well. The normalized heat flux remains close to unity for $Ro^{-1} \lesssim 2$, beyond which it decreases. This indicates that slow rotation does not affect the heat transport in the slender cell, in line with observations in wider convection cells for moderate Prandtl numbers~\citep{Wedi:JFM2021}. Figure~\ref{fig:Nu_Re_iRo} also shows that there is no enhancement of heat transport at moderate rotation rates, in contrast to that at large Prandtl numbers due to the so-called `Ekman pumping' mechanism~\citep{Stevens:PRL2009, Zhong:PRL2009, Zhong:JFM2010, Chong:PRL2017}, but the absence of heat flux enhancement in the slender cell data is consistent with low-$Pr$ RRBC in more extended domains~\citep{Zhong:PRL2009}.

The normalized Reynolds number $Re/Re_0$ with $Ro^{-1}$ is plotted in figure~\ref{fig:Nu_Re_iRo}(b). The trend is qualitatively similar to that of the normalized heat flux; weak rotation ($Ro^{-1} \lesssim 2$) does not affect momentum transport. For $Ro^{-1} \gtrsim 2$, the normalized momentum flux decreases but the data for $Ra = 10^8$ lies below those for $Ra \geq 10^9$ at higher $Ro^{-1}$. The suppression of the momentum transport is weaker than for heat transport; at $Ro^{-1} = 10$, the normalized Nusselt number $Nu/Nu_0 \approx 0.45$ whereas $Re/Re_0 \approx 0.8$, both for $Ra = 10^{10}$.

\section{Temperature gradient in the bulk region and viscous boundary layer}
\label{sec:dTdz}

\begin{figure}
\captionsetup{width=1\textwidth}
\centerline{\includegraphics[width=\textwidth]{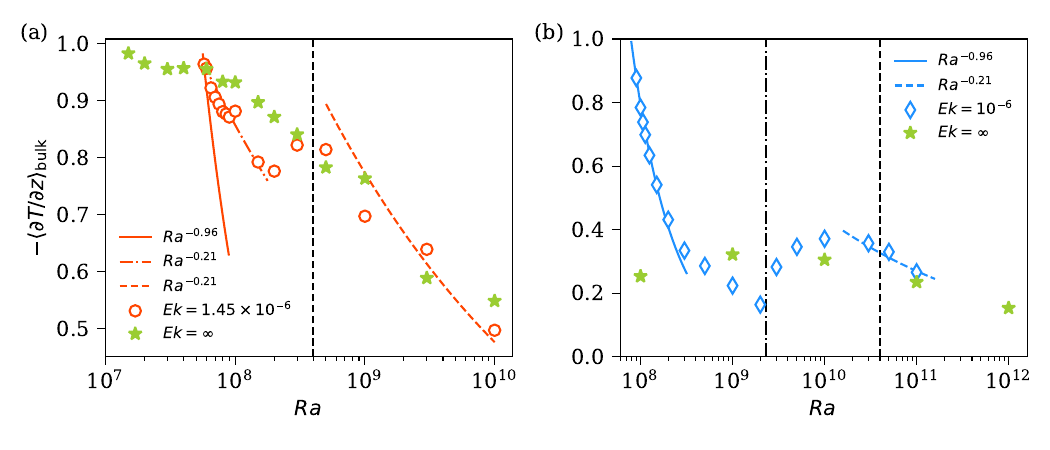}}
\caption{Mean vertical temperature gradient in the bulk region between $z = 0.25H$ and $z = 0.75H$ as a function of $Ra$ from non-rotating (green stars) and rotating (open symbols) slender cells for $Pr = 0.021$ (a) and $Pr = 1$ (b). Mean gradient decreases monotonically with $Ra$ in the non-rotating convection, whereas a non-monotonic trend is observed in the rotating convection. Solid and dashed curves are guides to the eye and not the best fits. Dash-dot vertical line in panel (b) indicates the transition $Ra \approx 23 Ek^{-4/3}$ between the cellular and plumes regimes, as found by \citet{Stellmach:PRL2014}. Dashed vertical lines in both panels correspond to $Ro = 0.2$. Non-rotating data in panel (b) is taken from \citet{Iyer:PNAS2020}.}
\label{fig:dTdz_Ra}
\end{figure}

The mean temperature in turbulent convection varies primarily in the thin thermal boundary layers (BLs) near the horizontal plates. However, severe lateral confinement causes temperature variation to be present also in the bulk region for moderate and low Prandtl numbers~\citep{Iyer:PNAS2020, Pandey:EPL2021, Pandey:PD2022}. The mean vertical temperature gradient $\partial T/\partial z$ decreases with increasing $Ra$ in the non-rotating case, whereas it changes in a specific manner in rotating convection~\citep{Julien:GAFD2012, Cheng:PRF2020, Guzman:PRF2022}. We compute mean vertical temperature gradient in the bulk region $\langle \partial T/\partial z \rangle_\mathrm{bulk}$ by performing average over the bulk volume with $z/H \in [0.25, 0.75]$ and plot it as a function of $Ra$ in figure~\ref{fig:dTdz_Ra}(a) for $Pr = 0.021$. The gradient remains close to unity and does not change significantly near the onset of non-rotating convection (green stars). Thus, bulk flow state in the vicinity of the onset does not differ much from the unmixed conduction state with $\partial T/\partial z = -1$. For $Ra > 10^8$, however, $-\langle \partial T/\partial z \rangle_\mathrm{bulk}$ decreases monotonically with $Ra$, but even for $Ra = 10^{10}$ low-$Pr$ RBC in the slender cell possesses a higher gradient than in the well-mixed case of extended domains.

The variation of $\langle \partial T/\partial z \rangle_\mathrm{bulk}$ in figure~\ref{fig:dTdz_Ra}(a) for rotating convection (red circles) is different. It has been known from the simulations of the asymptotically reduced equations~\citep{Sprague:JFM2006, Julien:GAFD2012} as well as direct numerical simulations of RRBC~\citep{Stellmach:PRL2014, Guzman:PRF2022} that the temperature gradient decreases steeply with $Ra$ in the cellular and columnar regimes, which occur in the vicinity of convective onset for moderate and large Prandtl numbers. With further increase of $Ra$, the gradient increases in the plumes regime and nearly saturates in the geostrophic regime, where the vertical coherence is lost~\citep{Stellmach:PRL2014}. For higher Rayleigh numbers in the rotation-affected regime, the gradient decreases again~\citep{Cheng:PRF2020}. Figure~\ref{fig:dTdz_Ra}(a) shows that the gradient for $Ek = 1.45 \times 10^{-6}$ decreases rapidly near the onset and starts to increase at $Ra = 2 \times 10^8$, before decreasing again for $Ra \geq 5 \times 10^8$. 

Figure~\ref{fig:dTdz_Ra}(b) shows the temperature gradient for $Pr = 1$ from the rotating case ($Ek = 10^{-6}$) and the non-rotating case (data taken from \citet{Iyer:PNAS2020}), both for slender cells. The bulk temperature gradient varies with $Ra$ qualitatively the same way as in the low-$Pr$ rotating convection. Near the critical Rayleigh number the gradient follows $Ra^{-0.96}$ scaling indicated by the blue solid curve. This scaling is consistent with the  onset results in simulations of asymptotic equations~\citep{Sprague:JFM2006, Julien:GAFD2012} as well as with the
DNS~\citep{King:JFM2013, Stellmach:PRL2014, Guzman:PRF2022} for moderate Prandtl numbers. With increasing $Ra$, the gradient decreases more slowly before increasing from $Ra = 2 \times 10^9$ up to $Ra \approx 6 \times 10^9$. As discussed earlier, an increasing gradient with $Ra$ is a characteristic of the plume region.  \cite{Stellmach:PRL2014} performed DNS in horizontally-periodic domain with both no-slip and free-slip plates and observed that the transition for $Pr = 1$ from the cellular to plumes region occurs at $Ra \approx 23 Ek^{-4/3}$ for the no-slip case. This corresponds to $Ra \approx 2.3 \times 10^9$ for the slender data; we indicate this transition $Ra$ by the dash-dot vertical line in figure~\ref{fig:dTdz_Ra}(b). It is interesting that the transition $Ra$ found for a horizontally-periodic domain identifies the transition for the slender data quite well.
It is observed in experiments~\citep{Cheng:PRF2020} and DNS~\citep{Guzman:PRF2022} that the temperature gradient decreases as $Ra^{-0.21}$ in the rotationally-affected regime, when the thermal forcing is significantly stronger than the critical value for the onset. The slender data at the largest Rayleigh numbers in figure~\ref{fig:dTdz_Ra}(b) nearly follow this scaling. Thus, the temperature gradient with $Ra$ in the rotating slender cell is qualitatively similar to that in wider cells, indicating again the dominance of rotation over confinement.

To see if the data in figure~\ref{fig:dTdz_Ra}(a) exhibit the scaling features just mentioned for moderate Prandtl numbers, we indicate the $Ra^{-0.96}$ scaling by a red solid curve, but find that the gradient near the onset decreases more slowly; instead, the data follow $\langle \partial T/\partial z \rangle_\mathrm{bulk} \sim Ra^{-0.21}$ scaling (red dash-dot curve). This is possibly an indication of a different flow state near the onset in low-$Pr$ convection. The $Ra^{-0.21}$ scaling in the rotation-affected regime is also indicated as a red dashed curve; we find from this exercise that the gradient for the largest $Ra$ in low-$Pr$ case is not very different. The dashed vertical lines in figure~\ref{fig:dTdz_Ra} correspond to $Ro = 0.2$, which suggests that the $Ra^{-0.21}$ scaling occurs when $Ro > 0.2$ and the rotational constraint in bulk region relaxes gradually.

\begin{figure}
\captionsetup{width=1\textwidth}
\centerline{\includegraphics[width=\textwidth]{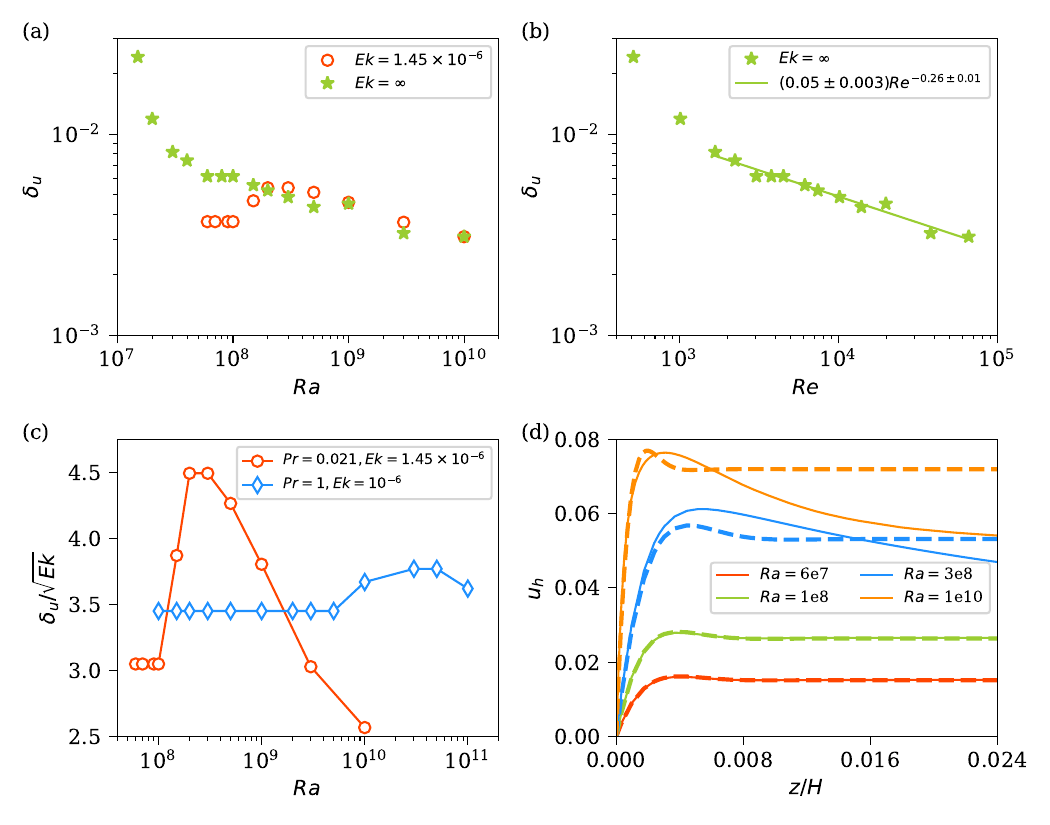}}
\caption{(a) Viscous boundary layer thickness for $Pr = 0.021$, averaged over both the horizontal plates, decreases with $Ra$ in non-rotating slender cell, whereas remains constant at low $Ra$ in rotating cell. (b) Thickness in the non-rotating cell as a function of $Re$. The best fit for $Re > 10^3$ shows that $\delta_u \sim Re^{-1/4}$. (c) Normalized Ekman layer thickness $\delta_u/\sqrt{Ek}$ remains a constant for a wider range of $Ra$ for $Pr = 1$ than for $Pr = 0.021$ simulations. (d) The horizontal velocity profile in the rotating slender cell (solid curves) for $Pr = 0.021$ follows the analytical Ekman layer profile (dashed curves) perfectly up to $Ra = 10^8$, but deviates for larger Rayleigh numbers.
}
\label{fig:delta_u}
\end{figure}

The thickness of the viscous boundary layer (VBL) near the horizontal plates decreases with increasing $Ra$ in non-rotating convection~\citep{King:JFM2013, Scheel:PRF2017, Bhattacharya:POF2018}. In rotating convection, however, the viscous boundary layer---also known as the Ekman layer---is controlled by the Ekman number; for weak thermal forcings, i.e., in rotationally-controlled regime, the Ekman layer thickness scales as $\sqrt{Ek}$~\citep{King:JFM2013, Guzman:PRF2022}. The VBL thickness $\delta_u$ is frequently determined using the rms horizontal velocity profile $u_h (z)$ where
$u_h = \sqrt{ \langle u_x^2 + u_y^2 \rangle_{A,t}}.$
Due to the imposed no-slip condition in our simulations, $u_h$ vanishes at the plates and increases rapidly as one moves away from them. We estimate $\delta_u$ as the distance of the first local maximum in the $u_h(z)$ profile from the horizontal plate. We compute $\delta_u$ at both the top and bottom plates and show the averaged thickness as a function of $Ra$ in figure~\ref{fig:delta_u}(a) for both the non-rotating and rotating slender cells. For the non-rotating case (green stars), $\delta_u$ decreases with $Ra$. In figure~\ref{fig:delta_u}(b), we plot the same data as a function of $Re$, which suggests that data for $Re > 10^3$ may be described by a single powerlaw. The best fit for this regime yields $\delta_u  = 0.05Re^{-0.26}$ scaling, which is in a qualitative agreement with the $Re^{-1/4}$ scaling observed in wider convection domains for moderate and high Prandtl numbers~\citep{King:JFM2013}.

The Ekman layer thickness for $Ek = 1.45 \times 10^{-6}$ in figure~\ref{fig:delta_u}(a) is nearly independent of $Ra$ for $Ra \leq 10^8$. The constancy of $\delta_u$ suggests that the VBL in this regime behaves as the classical Ekman layer, which results from the balance between the viscous and Coriolis forces~\citep{King:JFM2013}. Figure~\ref{fig:delta_u}(a) also reveals that a considerable variation in $\delta_u$ is observed for higher Rayleigh numbers. Further, the difference between the rotating and non-rotating data becomes very small for $Ra \geq 10^9$, which indicates the increasing dominance of thermal forcing over rotation as $Ra$ increases. To see the $Ek$-dependence of $\delta_u$, we plot the normalized thickness $\delta_u/\sqrt{Ek}$ as a function of $Ra$ in figure~\ref{fig:delta_u}(c) and also include the data from the $Pr = 1, Ek = 10^{-6}$ simulations. The figure shows that $\delta_u \sim \sqrt{Ek}$ scaling is indeed observed for both the Prandtl numbers at low thermal forcings and the prefactor $\approx 3$ for $Pr = 0.021$ and $\approx 3.5$ for $Pr = 1$ simulations. These prefactors are in the range of values reported in RRBC in wider domains~\citep{King:JFM2013, Guzman:PRF2022}. Figure~\ref{fig:delta_u}(c) also shows that the range of Rayleigh numbers over which the VBL is of the Ekman type is wider for $Pr = 1$ than for $Pr = 0.021$, which indicates the inertial nature of low-$Pr$ RRBC and consistent with the findings of \citet{Guzman:PRF2022}. 

We further probe the Ekman layer in the slender cell by investigating the form of the rms horizontal velocity profile $u_h(z)$ near the plates. For the classical Ekman layer above a no-slip plate, the velocity profile can be obtained analytically by considering a geostrophic  bulk flow, where the horizontal pressure gradients are balanced by the Coriolis forces and assuming that the same horizontal pressure gradients exist within the boundary layer region. Following \citet{Kundu:book} and \citet{Guzman:PRF2022}, we find that $u_h(z)$ near the plate can be described by
\begin{equation}
u_h(z) = U_h [1 - 2 \cos(z/\delta_U) {\rm e}^{-z/\delta_U} + {\rm e}^{-2z/\delta_U} ]^{1/2}. \label{eq:Ekman_profile}
\end{equation}
Here, $U_h = \sqrt{U_x^2+U_y^2}$ is the rms horizontal velocity in the geostrophic bulk, with $U_x$ and $U_y$ being the horizontal velocity components. The parameter $\delta_U$ corresponds to the thickness of the Ekman layer. In figure~\ref{fig:delta_u}(d), we show $u_h(z)$ for four Rayleigh numbers from the simulations at $Pr = 0.021, Ek = 1.45 \times 10^{-6}$ as solid curves. We fit these profiles using equation~\eqref{eq:Ekman_profile} and determine the parameters $U_h$ and $\delta_U$, and the resulting profiles obtained from equation~\eqref{eq:Ekman_profile} with the fitted parameters are exhibited as dashed curves in figure~\ref{fig:delta_u}(d). We observe that the profiles for $Ra \leq 10^8$ can be described excellently by the analytical profile~\eqref{eq:Ekman_profile}. However, deviation starts to appear for $Ra \geq 1.5 \times 10^8$. Figure~\ref{fig:delta_u}(d) exhibits that equation~\eqref{eq:Ekman_profile} still describes the near-wall profiles for all Rayleigh numbers. Thus, the VBL in the slender cell for $Pr = 0.021$ is of the Ekman type only up to $Ra = 10^8$, which is consistent with the inference from figure~\ref{fig:delta_u}(c). Note that similar results were reported in RRBC in horizontally-periodic boxes by \citet{Guzman:PRF2022}.

\section{Conclusions}
\label{sec:concl}

The center of attention in this paper is convection of low-$Pr$ fluids (chosen here to be 0.021) at a range of Rayleigh numbers up to $10^{10}$, with variable rotation rates. For comparison, we have also performed simulations for $Pr = 1$. By necessity, the aspect ratio is small. From a comparison of the present results with those for several different conditions, including convection in wider cells (where possible), we deduce a variety of results, a few of which are listed below.  

First, the flow structure, which is initially helical, develops progressively finer components with increasing thermal forcing. The flow structure feels the rotation strongly near the onset, with suppressed variation along the vertical direction. With increasing $Ra$, however, the resilience increases and the flow configuration for $Ra = 10^{10}$ (figure~\ref{fig:iso_Ek_fixed}(d,h)) shows strong resemblance with its non-rotating counterpart in figure~\ref{fig:iso_nonrot}. In spite of this feature, the essentially isothermal bulk region, observed to exist in wider convection domains, is absent in the slender cell. Yet, the heat transport scaling is the same as in wider cells for a given high Rayleigh number, which shows the secondary role of the bulk flow for global heat transport.

We found that near the onset, the supercritical behavior is qualitatively independent of $Pr$. For intermediate $Ra$, the Nusselt number in the non-rotating slender cell increases steeply with $Ra$; we found $Nu-1 \sim Ra$ for $6 \times 10^7 \leq Ra \leq 5 \times 10^8$. This increase is steeper than those in convection domains of $\Gamma \geq 0.5$, where $Nu \sim Ra^{\beta}$ with $\beta \in [0.25-0.30]$ have been observed~\citep{Glazier:Nature1999, Cioni:JFM1997, Scheel:PRF2017, Schindler:PRL2022}.

We found that $Nu-1$ in the rotating cell increases approximately as $Ra^{1.3}$ for the intermediate Rayleigh numbers, which is not very different from the $Nu-1 \sim Ra^{3/2}$ scaling proposed for the geostrophic regime~\citep{Julien:PRL2012}. Further, we observed $Nu \sim Ra^{0.95}$ scaling for $10^8 \leq Ra \leq 10^9$, which is close to that found in a wider cell at a similar Prandtl number~\citep{Aurnou:JFM2018}. For $Ra \geq 10^{9}$ the $Nu$ data agree reasonably well with the canonical $Ra^{1/3}$ scaling observed in non-rotating wider convection cells~\citep{Niemela:Nature2000}. We also studied the effects of increasing rotation on the integral transports and the flow structure for fixed thermal forcings and observed that these flow properties in the slender cell are altered in very similar manners to those reported in $\Gamma \geq 0.5$ rotating flows.

We obtained the mean temperature gradient in the bulk region of the rotating slender cells for $Pr = 0.021$ and $Pr = 1$ and found that its variation with $Ra$ is similar to those reported in extended domains. We also analysed the width of the Ekman layer and the velocity profile in the region near the plate and observed that they exhibit very similar behavior observed in rapidly rotating convective flows in wider domains. Thus, the effects of rotation on the slender convection are similar to those in extended convection, even though the non-rotating case exhibits differing behavior, as long as $Ra$ is high enough. 

We point out that the maximum value of the convective supercriticality, $Ra/Ra_c$, explored in the present work for $Ek = 1.45 \times 10^{-6}$ is nearly 200. This value is not very large for the non-rotating convection. In RRBC, however, the flow characteristics change rapidly as $Ra/Ra_c$ increases from unity and one observes richer dynamical regimes compared to those in non-rotating convection over a relatively shorter range of $Ra/Ra_c$. In addition to our own observations, we cite~\citet{Julien:GAFD2012, Guzman:PRF2022} and \citet{Ecke:ARFM2023} also for supporting evidence.

Our study, which is based on simulations in a slender cell of a fixed aspect ratio 0.1, suggests that rotation influences convection more strongly than the geometric confinement. This is an important conclusion, as the rotating convective flows could be explored at higher Rayleigh numbers using slender domains, opening new parameter ranges not accessible to wider convection cells. We reiterate that, while with decreasing $\Gamma$ the sidewall boundary layer is expected to have an increasingly stronger influence on the dynamics of RRBC, the rotation effects often overwhelm other factors. It is, of course, obvious that further studies with varying $\Gamma$ would help us better understand the interplay between the effects of rotation and confinement.

\backsection[Supplementary data]{\label{SupMat}Supplementary material and movies are available at \\https://doi.org/10.1017/jfm.2024...}

\backsection[Acknowledgements]{We thank J\"org Schumacher for useful input on the work. We also acknowledge valuable discussions with Ravi Samtaney, who sadly passed away while the work was in progress. The authors gratefully acknowledge {\sc Shaheen II} of KAUST, Saudi Arabia (under Project Nos. k1491 and k1624) and {\sc Dalma} and {\sc Jubail} clusters at NYU Abu Dhabi for providing computational resources. 
}

\backsection[Funding]{This material is based upon work supported by Tamkeen under the NYU Abu Dhabi Research Institute grant G1502 and by the KAUST Office of Sponsored Research under Award URF/1/4342-01.}

\backsection[Declaration of Interests]{The authors report no conflict of interest.}

\backsection[Data availability statement]{The data that support the findings of this study are available from the corresponding author upon reasonable request.}

\backsection[Author ORCIDs]{\\
A. Pandey, https://orcid.org/0000-0001-8232-6626;\\
K. R. Sreenivasan, https://orcid.org/0000-0002-3943-6827}

\appendix

\section{Simulation parameters}
\label{sec:app_sim}

\begin{table}
\captionsetup{width=1\textwidth}
  \begin{center}
\def~{\hphantom{0}}
  \begin{tabular}{lcccccccc}
$Ra$ & $N_e \times N^3$ & $Nu$ & $Re$ & $t_\mathrm{sim} \, (t_f)$ & $\Delta_z/\eta$ \\
$1.5 \times 10^7$ & $192000 \times 3^3$ & $1.05 \pm 0.001$ & $512 \pm 1$ & 2547 & 0.32 \\
$2 \times 10^7$ & $192000 \times 3^3$ & $1.19 \pm 0.001$ & $1010 \pm 1$ & 1448 & 0.47 \\ 
$3 \times 10^7$ & $192000 \times 3^3$ & $1.43 \pm 0.004$ & $1673 \pm 1$ & 1255 & 0.65 \\ 
$4 \times 10^7$ & $192000 \times 3^3$ & $1.67 \pm 0.07$ & $2230 \pm 13$ & 950 & 0.78 \\ 
$6 \times 10^7$ & $192000 \times 3^3$ & $2.03 \pm 0.07$ & $3037 \pm 6$ & 997 & 0.97 \\ 
$8 \times 10^7$ & $192000 \times 3^3$ & $2.40 \pm 0.01$ & $3775 \pm 8$ & 995 & 1.12 \\ 
$1 \times 10^8$ & $192000 \times 3^3$ & $2.81 \pm 0.001$ & $4496 \pm 4$ & 929 & 1.26 \\ 
$1.5 \times 10^8$ & $192000 \times 5^3$ & $3.84 \pm 0.1$ & $6125 \pm 11$ & 269 & 1.00 \\ 
$2 \times 10^8$ & $192000 \times 5^3$ & $4.64 \pm 0.09$ & $7400 \pm 15$ & 385 & 1.14 \\ 
$3 \times 10^8$ & $192000 \times 7^3$ & $6.84 \pm 0.42$ & $10114 \pm 30$ & 140 & 1.04 \\ 
$5 \times 10^8$ & $192000 \times 7^3$ & $9.60 \pm 0.07$ & $13896 \pm 3$ & 110 & 1.30 \\ 
$1 \times 10^9$ & $537600 \times 7^3$ & $12.9 \pm 0.63$ & $19800 \pm 55$ & 61.5 & 1.24 \\ 
$3 \times 10^9$ & $537600 \times 7^3$ & $27.2 \pm 0.66$ & $37905 \pm 122$ & 36.9 & 2.00 \\ 
$1 \times 10^{10}$ & $537600 \times 13^3$ & $39.9 \pm 8.5$ & $65715 \pm 804$ & 29.2 & 1.69
  \end{tabular}
  \caption{Parameters of DNS for $Pr = 0.021$ in the non-rotating cylindrical cell of $\Gamma = 0.1$: the number of mesh cells $N_e \times N^3$ in the entire flow domain, where $N_e$ is the number of elements and $N$ the polynomial order of the Lagrangian interpolation; $Nu$ is the globally-averaged heat transport estimated using \eqref{eq:Nu_uzT} and $Re$ is the Reynolds number based on the root-mean-square velocity. Integration time in free-fall units in the statistically steady state is represented by $t_\mathrm{sim}$, and the maximum value of ratio of the local vertical grid spacing $\Delta_z(z)$ to the local Kolmogorov scale $\eta(z)$ is shown in the last column. Error bars indicate the difference in the mean values of the two-halves of the data sets.}
  \label{table:details_nonrot}
  \end{center}
\end{table}

\begin{table}
\captionsetup{width=1\textwidth}
  \begin{center}
\def~{\hphantom{0}}
  \begin{tabular}{lcccccccc}
$Ra$ & $N_e \times N^3$ & $Nu$ & $Re$ & $t_\mathrm{sim} \, (t_f)$ & $\Delta_z/\eta$ \\
$5.75 \times 10^7$ & $192000 \times 3^3$ & $1.061 \pm 0.001$ & $737 \pm 1$ & 1947 & 0.47 \\
$6.0 \times 10^7$ & $192000 \times 3^3$ & $1.073 \pm 0.001$ & $818 \pm 1$ & 1890 & 0.50 \\ 
$6.5 \times 10^7$ & $192000 \times 3^3$ & $1.133 \pm 0.002$ & $1167 \pm 1$ & 1434 & 0.59 \\ 
$7.0 \times 10^7$ & $192000 \times 3^3$ & $1.167 \pm 0.001$ & $1360 \pm 1$ & 2301 & 0.63 \\ 
$7.5 \times 10^7$ & $192000 \times 3^3$ & $1.189 \pm 0.005$ & $1510 \pm 1$ & 2579 & 0.69 \\ 
$8.0 \times 10^7$ & $192000 \times 3^3$ & $1.215 \pm 0.001$ & $1629 \pm 1$ & 2475 & 0.72 \\ 
$8.5 \times 10^7$ & $192000 \times 3^3$ & $1.226 \pm 0.003$ & $1684 \pm 1$ & 2423 & 0.73 \\ 
$9.0 \times 10^7$ & $192000 \times 3^3$ & $1.240 \pm 0.007$ & $1786 \pm 1$ & 2390 & 0.74 \\
$1.0 \times 10^8$ & $192000 \times 3^3$ & $1.254 \pm 0.012$ & $1885 \pm 4$ & 1288 & 0.77 \\ 
$1.5 \times 10^8$ & $192000 \times 3^3$ & $1.80 \pm 0.03$ & $3719 \pm 9$ & 454 & 1.13 \\ 
$2 \times 10^8$ & $192000 \times 3^3$ & $2.39 \pm 0.02$ & $5276 \pm 9$ & 355 & 1.42 \\ 
$3 \times 10^8$ & $192000 \times 5^3$ & $3.24 \pm 0.07$ & $7121 \pm 20$ & 190 & 1.11 \\ 
$5 \times 10^8$ & $192000 \times 7^3$ & $5.62 \pm 0.32$ & $11109 \pm 51$ & 125 & 1.11 \\ 
$1 \times 10^9$ & $537600 \times 7^3$ & $11.2 \pm 0.32$ & $18595 \pm 81$ & 63.0 & 1.20 \\ 
$3 \times 10^9$ & $537600 \times 7^3$ & $22.1 \pm 2.8$ & $34989 \pm 330$ & 54.2 & 1.90 \\ 
$1 \times 10^{10}$ & $537600 \times 13^3$ & $40.6 \pm 9.0$ & $65306 \pm 742$ & 30.6 & 1.66
  \end{tabular}
  \caption{The same DNS parameters for $Pr = 0.021$ in a rapidly rotating cylindrical cell of $\Gamma = 0.1$ for $Ek = 1.45 \times 10^{-6}$.}
  \label{table:details_Ek1e-6}
  \end{center}
\end{table}

\begin{table}
\captionsetup{width=1\textwidth}
  \begin{center}
\def~{\hphantom{0}}
  \begin{tabular}{lccccccccc}
$Ra$ & $Ro^{-1}$ & $N_e \times N^3$ & $Nu$ & $Re$ & $t_\mathrm{sim} \, (t_f)$ & $\Delta_z/\eta$ \\
$10^8$ & 0 & $192000 \times 3^3$ & $2.81 \pm 0.001$ & $4496 \pm 4$ & 929 & 1.26 \\
$10^8$ & 0.25 & $192000 \times 3^3$ & $2.82 \pm 0.13$ & $4515 \pm 17$ & 583 & 1.27 \\
$10^8$ & 0.50 & $192000 \times 3^3$ & $2.76 \pm 0.02$ & $4435 \pm 1$ & 583 & 1.25 \\
$10^8$ & 1 & $192000 \times 3^3$ & $2.72 \pm 0.17$ & $4391 \pm 16$ & 583 & 1.25 \\
$10^8$ & 2 & $192000 \times 3^3$ & $2.53 \pm 0.02$ & $4187 \pm 10$ & 706 & 1.21 \\
$10^8$ & 5 & $192000 \times 3^3$ & $1.79 \pm 0.08$ & $3172 \pm 8$ & 723 & 1.02 \\
$10^8$ & 10 & $192000 \times 3^3$ & $1.25 \pm 0.01$ & $1885 \pm 4$ & 1288 & 0.77 \\
$10^9$ & 0 & $537600 \times 7^3$ & $12.9 \pm 0.72$ & $19823 \pm 60$ & 61.8 & 1.24 \\
$10^9$ & 1 & $537600 \times 7^3$ & $12.3 \pm 0.29$ & $19310 \pm 3$ & 39.7 & 1.23 \\
$10^9$ & 2 & $537600 \times 7^3$ & $11.6 \pm 0.12$ & $18800 \pm 16$ & 45.3 & 1.21 \\
$10^9$ & 3 & $537600 \times 7^3$ & $10.5 \pm 0.84$ & $18104 \pm 206$ & 45.0 & 1.18 \\
$10^9$ & 5 & $537600 \times 5^3$ & $8.34 \pm 1.2$ & $16734 \pm 146$ & 90.8 & 1.50 \\
$10^9$ & 10 & $537600 \times 5^3$ & $3.92 \pm 0.43$ & $12359 \pm 78$ & 125 & 1.20 \\
$10^9$ & 20 & $537600 \times 5^3$ & $2.15 \pm 0.08$ & $7928 \pm 40$ & 136 & 0.95 \\
$10^{10}$ & 0 & $537600 \times 13^3$ & $39.9 \pm 8.5$ & $65715 \pm 804$ & 29.2 & 1.69 \\
$10^{10}$ & 1 & $537600 \times 13^3$ & $40.6 \pm 9.0$ & $65306 \pm 742$ & 30.6 & 1.66 \\
$10^{10}$ & 2 & $537600 \times 13^3$ & $34.8 \pm 2.3$ & $61958 \pm 202$ & 24.2 & 1.62 \\
$10^{10}$ & 5 & $537600 \times 13^3$ & $31.5 \pm 5.2$ & $60277 \pm 1114$ & 27.9 & 1.61 \\
$10^{10}$ & 10 & $537600 \times 11^3$ & $17.8 \pm 4.8$ & $50783 \pm 595$ & 26.8 & 1.61 \\
$10^{10}$ & 20 & $537600 \times 9^3$ & $5.73 \pm 1.5$ & $31880 \pm 484$ & 88.4 & 1.33 \\
$10^{10}$ & 30 & $537600 \times 7^3$ & $3.88 \pm 0.30$ & $24455 \pm 58$ & 303 & 1.59 \\
  \end{tabular}
  \caption{Parameters of DNS for $Pr = 0.021$ with varying rotation frequency.}
  \label{table:details_Ra1e10}
  \end{center}
\end{table}

\begin{figure}
\captionsetup{width=1\textwidth}
\centerline{\includegraphics[width=\textwidth]{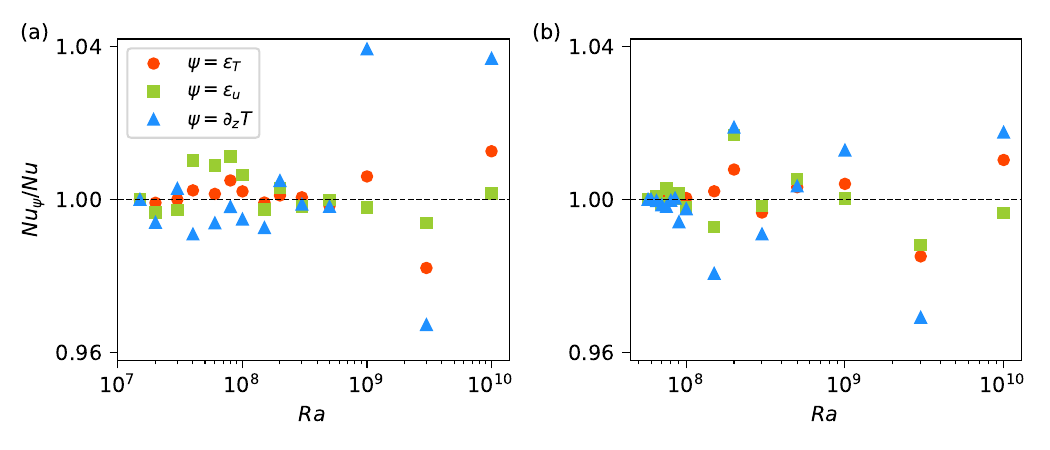}}
\caption{Nusselt numbers computed using the thermal and viscous dissipation rates and the wall temperature gradient agree with $Nu$ computed from \eqref{eq:Nu_uzT} within 4\% for all simulations. Panel (a) shows the ratios $Nu_{\varepsilon_T}/Nu$, $Nu_{\varepsilon_u}/Nu$, and $Nu_{\partial_z T}/Nu$ for $Pr = 0.021$ and $Ek = \infty$, whereas data from the rotating cell for $Ek = 1.45 \times 10^{-6}$ are shown in panel (b).}
\label{fig:Nu_various}
\end{figure}

We collect important parameters of DNS in the non-rotating and rotating slender cells in tables~\ref{table:details_nonrot} and \ref{table:details_Ek1e-6}, respectively. Table~\ref{table:details_Ra1e10} contains relevant parameters of simulations for fixed Rayleigh numbers and varying rotation rates. In addition to comparing the smallest grid spacing with the Kolmogorov length scale (see \S\ref{sec:numerical}), we examine the convergence of the heat flux using different methods~\citep{Pandey:PD2022}; a properly resolved simulation should yield the same global heat transport when computed from different approaches. The exact relations of RBC link the volume and time averaged thermal and kinetic energy dissipation rates with the Nusselt number~\citep{Shraiman:PRA1990} and the heat fluxes from the energy and the thermal dissipation rates are estimated as
\begin{eqnarray}
Nu_{\varepsilon_u} & = & 1 + \frac{H^4}{\nu^3} \frac{Pr^2}{Ra} \langle \varepsilon_u \rangle_{V,t}, \\
Nu_{\varepsilon_T} & = & \frac{H^2}{\kappa (\Delta T)^2} \langle \varepsilon_T \rangle_{V,t}.
\end{eqnarray}
At the horizontal plates, the heat is entirely transported due to molecular diffusion and the area-averaged flux at the plates is estimated using the vertical temperature gradient as
\begin{equation}
Nu_{\partial_zT} = - \frac{H} {\Delta T} \left \langle \left( \frac{\partial T}{\partial z} \right)_{z=0,H}  \right \rangle_{A,t} .
\end{equation}
We plot the ratios $Nu_{\varepsilon_T}/Nu$, $Nu_{\varepsilon_u}/Nu$, and $Nu_{\partial_z T}/Nu$ in figure~\ref{fig:Nu_various}(a) for the non-rotating simulations and in figure~\ref{fig:Nu_various}(b) for simulations at $Ek = 1.45 \times 10^{-6}$. The ratios depart from unity by a maximum of 4\% for all the simulations, affirming that the simulations are adequately resolved.


\end{document}